# Far-Ultraviolet Observations of Outflows from IR-Luminous Galaxies

Claus Leitherer

*Space Telescope Science Institute[1], 3700 San Martin Drive, Baltimore, MD 21218*

leitherer@stsci.edu

Rupali Chandar

*The University of Toledo, Dept. of Physics & Astronomy, Toledo, OH 43606*

Rupali.Chandar@utoledo.edu

Christy A. Tremonti

*Univ. of Wisconsin-Madison, Dept. of Astronomy, 475 N. Charter St., Madison, WI 53706*

tremonti@astro.wisc.edu

Aida Wofford

*Space Telescope Science Institute[1], 3700 San Martin Drive, Baltimore, MD 21218*

wofford@stsci.edu

Daniel Schaerer

*Observatoire de Genève, 51 chemin des Maillettes, CH 1290 Versoix, Switzerland*

daniel.schaerer@unige.ch

---

[1] Operated by the Association of Universities for Research in Astronomy, Inc., under NASA contract NAS5−26555

**Abstract**

We obtained medium-resolution ultraviolet (UV) spectra between 1150 and 1450 Å of the four UV-bright, infrared (IR)-luminous starburst galaxies IRAS F08339+6517, NGC 3256, NGC 6090, and NGC 7552 using the Cosmic Origins Spectrograph onboard the Hubble Space Telescope. The selected sightlines towards the starburst nuclei probe the properties of the recently formed massive stars and the physical conditions in the starburst-driven galactic superwinds. Despite being metal-rich and dusty, all four galaxies are strong Lyman-α emitters with equivalent widths ranging between 2 and 13 Å. The UV spectra show strong P Cygni-type high-ionization features indicative of stellar winds and blueshifted low-ionization lines formed in the interstellar and circumgalactic medium. We detect outflowing gas with bulk velocities of ~400 km s$^{-1}$ and maximum velocities of almost 900 km s$^{-1}$. These are among the highest values found in the local universe and comparable to outflow velocities found in luminous Lyman-break galaxies at intermediate and high redshift. The outflow velocities are unlikely to be high enough to cause escape of material from the galactic gravitational potential. However, the winds are significant for the evolution of the galaxies by transporting heavy elements from the starburst nuclei and enriching the galaxy halos. The derived mass outflow rates of ~100 M$_\odot$yr$^{-1}$ are comparable to, or even higher than the star-formation rates. The outflows can quench star formation and ultimately regulate the starburst as has been suggested for high-redshift galaxies.

*Key words:* Ultraviolet: galaxies − Galaxies: halos − Galaxies: ISM − Galaxies: starburst − Galaxies: star formation

# 1. Introduction

The absorption spectra of quasi-stellar objects (QSOs) demonstrate conclusively that metals in the universe are found far from their production sites. Mg II absorbers, for example, are associated with normal galaxies at impact parameters of up to tens of kpc, well outside their luminous dimensions. Historically, Mg II absorbers were the first systems used for the study of QSO absorber-galaxy relationships (Bergeron 1986). This choice was dictated by practical considerations: among all strong ions in the intergalactic medium (IGM) only Mg II was accessible from the ground at low redshift *(z < 2)*. Since then, advances in ground- and space-based telescopes and detectors have pushed the discovery space to lower and higher *z*, as well as to numerous other ions (e.g., Adelberger et al. 2005; Ménard et al. 2011). Most recently, Tumlinson et al. (2011) took advantage of the high ultraviolet (UV) sensitivity of the Cosmic Origins Spectrograph (COS) onboard the Hubble Space Telescope (HST) to observe the 1032,38 Å doublet of O VI, the most accessible tracer of hot and/or highly ionized circumgalactic gas at *z < 0.5*. However, metals are not only found in the immediate vicinity of galaxies but also in the Lyman-α forest, which traces large-scale structures of moderate overdensity relative to the cosmic mean (Pettini 2004).

On the theoretical side there remain serious difficulties in understanding how metals could be transported between such large volumes; this problem applies to all scales, from galactic[2] halos (e.g., Charlton & Churchill 1998) to Mpc-size Lyman-α clouds (e.g., Gnedin 1998). Various models have been proposed in the literature: outflows (Bond et al. 2001), infalling gas (Tinker & Chen 2008), or extended disks (Steidel et al. 2002), but consensus has not yet emerged. Observationally, no compelling correlations between absorber properties and those of their associated galaxies have been found. In particular, the physical processes governing the observed equivalent width, the basic observable quantity, and the origin of its distribution function are not yet understood.

An alternative technique of probing the circumgalactic gas is to directly target galaxies and measure the self-absorption in their spectra. Such an analysis does not provide any information on the distance between the absorbing gas and the galaxy, and the connection to intervening absorber systems is less straightforward. However, the galaxy properties are much better constrained in this case, as opposed to the quasar absorption-line technique, where the presence of an intervening disk galaxy is often merely inferred. Nevertheless, there is general agreement on a strong link between these two methods of detecting gas in and around galaxies. Absorption-line studies of circumgalactic gas have been performed at low (e.g., Heckman et al. 2000; Martin 2005; Rupke & Veilleux 2005; Rupke, Veilleux, & Sanders 2005a,b,c; Grimes et al. 2009; Chen et al. 2010), intermediate (e.g., Tremonti et al. 2007; Weiner et al. 2009; Coil et al. 2011; Kornei et al. 2012), and high (e.g., Pettini et al. 2002; Steidel et al. 2010; Jones, Stark, & Ellis 2012) redshift. Owing to the wavelengths of the diagnostic absorptions lines, observations of high-$z$ galaxies are done from the ground, whereas space-UV observations are required for nearby galaxies, except for the studies employing the optical Na D doublet.

---

[2] Throughout this paper "galactic" refers to galaxies other than the Milky Way, whereas "Galactic" is a reference to the Milky Way.

Star-forming and starburst galaxies are of particular interest in this context. It is well-established that galactic-scale outflows of gas are a ubiquitous phenomenon in the most actively star-forming galaxies in the local universe (Chen et al. 2010) and at high redshift (Shapley et al. 2003). These outflows are potentially very important in the evolution of galaxies and the IGM (Veilleux, Cecil, & Bland-Hawthorn 2005). For example, by selectively blowing metals out of shallow galactic potential wells, they may explain the tight relation between the mass and metallicity in galaxies (Tremonti et al. 2004). This same process would have enriched and heated the IGM in metals at early times (e.g. Adelberger et al. 2003), and explain why the majority of metals in galaxy clusters are in the intracluster medium (e.g., Loewenstein 2006). At the same time, *local* starbursts deserve detailed studies in their own right. They allow us to study star formation in extreme environments, the heating and cooling of the interstellar medium (ISM), and the dynamical effects of supernovae and winds. Their cosmological relevance has been highlighted by their many similarities to star forming galaxies at high-redshift. In particular, local UV-bright starbursts appear to be good analogs to Lyman-break galaxies (Shapley et al. 2003; Pettini 2008; Heckman et al. 2011) and so can be used as a training set to provide a thorough understanding of rest-frame UV spectral diagnostics that are critical for studying star-formation in the early universe. Starbursts can contain millions of OB stars, and hence they also offer a unique opportunity to test theories of the evolution of massive stars.

The UV spectral window is a key for understanding starbursts. This is where the *intrinsic* energy distribution of a starburst stellar population peaks. These hot massive stars have photospheric and wind spectral features in the UV that provide key information about the metallicity, initial mass function (IMF) and starburst age (Tremonti et al. 2001; Chandar et al. 2003; Wofford, Leitherer, & Chandar 2010; Leitherer 2011). The UV region contains powerful (and in some cases, unique) diagnostics of the physical, chemical, and dynamical properties of the ISM from cold molecular through hot coronal phases. Previous UV spectroscopy with HST or other satellites has focused almost exclusively on blue, and therefore metal-poor, low-luminosity (and, by implication, low-mass) starburst galaxies but has neglected the powerful, infrared-(IR) bright (but UV-faint) nuclear starbursts. Yet, the more massive IR-bright starbursts may have very different population properties in terms of the IMF, specific star-formation intensity, and their different halo masses may affect their relation to the circumgalactic gas (Kaviraj 2009; Howell et al. 2010; Kennicutt & Evans 2012).

With COS we finally have a spectrograph whose superior sensitivity and resolving power allow us to study these objects in unprecedented detail and at the same time requiring a moderate number of HST orbits. We have obtained COS G130M observations of a sample of four IR-luminous starburst galaxies whose proximity and UV brightness make them excellent targets. At the same time, these galaxies are ideally suited for extrapolation to the properties of starburst galaxies at cosmological distances.

In Section 2 we introduce the galaxy sample and compile the relevant properties from the literature. The observational strategy and the reduction pipeline are discussed in Section 3. In Section 4 we describe the spectral properties related to massive stars and the determination of the dust reddening and the stellar content. A discussion of the Lyman-α line is in Section 5. Section 6 describes the properties of the interstellar and circumgalactic gas. Finally, Section 7 presents our conclusions.

## 2. Selection and Properties of the Program Galaxies

We searched the Mikulski Archive for Space Telescopes (MAST) at STScI and identified four IR-luminous ($L_{IR} > 10^{11}$ L$_\odot$) starburst galaxies with existing low-resolution UV spectroscopy whose UV brightness permits cost-effective spectroscopy with COS. Naively, one would expect dusty, IR-luminous galaxies to be dauntingly faint in the UV. However, our understanding of the geometrical distribution of dust in starbursts and its effects on the emergent UV spectrum is still incomplete. The available data strongly suggest that (quite surprisingly) much of the dust responsible for the UV extinction is apparently distributed around the starburst in the form of a moderately inhomogeneous foreground screen or sheath surrounding the starburst (Calzetti 2001, 2009). This empirical result has the fortunate implication that (with the exception of the most heavily dust-shrouded starbursts like M 82 or Arp 220) we are able to directly study much of the stellar population responsible for powering the starburst via observations in the UV (albeit often through several magnitudes of attenuation). The four selected galaxies are IRAS F08339+6517, IRAS F10257−4339 (= NGC 3256), IRAS F16104+5235 (= NGC 6090), and IRAS F23133−4251 (= NGC 7552). In the following we refer to the galaxies as F08339+6517, NGC 3256, NGC 6090, and NGC 7552. All four objects are included in the Great Observatories All-sky LIRG Survey (GOALS), which combines imaging and spectroscopic data of ~180 Luminous Infrared Galaxies (LIRGs) from the Spitzer, HST, Chandra and GALEX space-borne observatories (Armus et al. 2009; Howell et al. 2010). We will make use of some of the results of GOALS in the present study. No strong AGN activity has been detected in any of the galaxies, except for weak LINER evidence in NGC 7552 (see below).

The selected sample was chosen to be used for a pilot project with COS. Its properties were consciously biased towards maximizing detection of outflows in the UV. Specifically, the galaxies are observed face-on, and they are known to have sight lines with low reddening. Observations of nearby dwarf starburst galaxies such as, e.g., NGC 4214 have demonstrated how the combined effects of supernovae and stellar winds can sweep the interstellar gas from the inner regions of young star clusters, thereby creating cavities and shells. The cavity interior, with the star cluster in its center has lower reddening and decreased gas density (Úbeda, Máiz-Apellániz, & MacKenty 2007). Therefore sight lines with low reddening may indicate the action of high-velocity outflows, and there may be a causal connection between high UV photon escape and the detection of winds.

A summary of some basic properties of the four galaxies collected from the literature is in Table 1. Columns 1 and 2 of this table list the names and alternate designations of the galaxies. In column 3 we give the morphological types, taken from the NASA Extragalactic Database (NED) or the Revised Shapley-Ames Catalog of Bright Galaxies (Sandage & Tammann 1987). The Milky Way foreground reddenings *E(B−V)*$_{MW}$ of Schlegel, Finkbeiner, & Davis (1998) as listed on NED are in column 4. In column 5 we give the heliocentric velocities ($v_{helio}$), also taken from NED. The values are significantly above the values expected for Milky Way lines, including for those of high-velocity clouds. Therefore blending of the intrinsic spectral features with Galactic lines is not an issue. The luminosity distances *D* (column 6) were taken from the compilation of Armus et al. (2009). They were derived by correcting $v_{helio}$ for the 3-attractor flow model of Mould et al. (2000) and adopting cosmological parameters $H_0 = 70$ km s$^{-1}$ Mpc$^{-1}$, $\Omega_{\text{Vacuum}} = 0.72$, and $\Omega_{\text{Matter}} = 0.28$ based on the five-year WMAP results (Hinshaw et al. 2009), as provided by NED. We calculated the projected linear size of the COS aperture (diameter 2.5″) at the distance of each galaxy ($A_{\text{COS}}$, column 7). In column 8 the total IR luminosities are tabulated. The values were again taken from Armus et al. who adopted the flux densities reported in the IRAS Revised Bright Galaxy Sample (Sanders et al. 2003) and the luminosity distances in column 6, together with the relation

$$L_{\text{IR}}/L_\odot = 4\pi D^2 \, F_{\text{IR}}/3.826 \times 10^{26}. \tag{1}$$

$F_{\text{IR}}$ can be determined from the IRAS 12, 25, 60 and 100 μm flux densities $f_{12\mu m}$, $f_{25\mu m}$, $f_{60\mu m}$, and $f_{100\mu m}$, respectively, using the formula given by Sanders & Mirabel (1996):

$$F_{\text{IR}} = 1.8 \times 10^{14} \, (13.48 f_{12\mu m} + 5.16 f_{25\mu m} + 2.58 f_{60\mu m} + f_{100\mu m}). \tag{2}$$

All units on the right-hand sides of equations (1) and (2) are in the mks system. In column 9 of Table 1 we report the oxygen abundances as published by the indicated references. Given the high oxygen abundances, these values should be viewed with care since abundance determinations in this parameter regime are quite uncertain (Kennicutt, Bresolin, & Garnett 2003). In the following we provide brief descriptions of the individual galaxies.

F08339+6517 is at 86 Mpc and was first reported in the IRAS Point Source Catalog. Margon et al. (1988) performed a multi-wavelength study and described the object as an exceptionally bright and compact starburst nucleus. Wiklind (1989) found abundant molecular gas from CO observations in this galaxy, concluding that it is a genuine starburst system. Cannon et al. (2004) presented VLA H I imaging of F08339+6517. They found extended neutral gas between this galaxy and a nearby neighbor, suggesting that interactions have played an important role in triggering the massive starburst in the primary galaxies. López-Sánchez, Esteban, & García-Rojas (2006) obtained optical imaging which suggests that the H I tail has been mainly formed from material stripped from the main galaxy. Their spectroscopy revealed weak Wolf-Rayet features in the central knot of the galaxy, consistent with ongoing star formation.

NGC 3256 with a total luminosity of log $L_{IR}/L_\odot$ = 11.6 is the most luminous galaxy in the local universe within $z < 0.01$ (Sargent, Sanders, & Phillips 1989). NGC 3256 is in the late stage of galaxy merging. A pair of long tidal tails extending 40 kpc on each side suggests a prograde-prograde merger of two gas-rich spiral galaxies of similar size. It also suggests that we view the system with a low inclination angle (English et al. 2003; Sakamoto, Ho, & Peck 2006). The center of the merging galaxy hosts a starburst seen at all observed wavelengths. High-resolution HST imaging revealed hundreds of bright young clusters near the galactic center (Alonso-Herrero et al. 2002). There are two central nuclei observed in X-rays, separated by ~5″. The two nuclei, along with the diffuse X-ray emission, support the scenario that NGC 3256 is powered by a starburst rather than an AGN (Lira et al. 2002).

NGC 6090 is an IR-luminous galaxy-galaxy merger at a distance of 137 Mpc, which is viewed face-on (Dinshaw et al. 1999). At optical wavelengths, NGC 6090 appears as a double nucleus system. The separation of the nuclei, as measured at radio wavelengths, is ~3.5 kpc. Arribas et al. (2004) found an irregular morphology in the innermost regions around the nuclei (1–2 kpc), where a "bridge" connects the northern part of these two regions. A common envelope extends up to about 7 kpc. Two large tidal tails extend at least 65 kpc to the south and 50 kpc northeast. There is considerable evidence for starburst activity, but no evidence at optical or radio wavelengths for a compact AGN. Sugai et al. (2004) have found a kiloparsec-scale region where active star formation is currently increasing the fraction of heavy elements.

NGC 7552 is part of the Grus quartet in the southern hemisphere at a distance of 24 Mpc. The galaxy is nearly face-on with an inclination of ~28° (Feinstein et al. 1990). NGC 7552 has been classified as a LINER by Durret & Bergeron (1988), based on their detection of weak [O I] λ6300. Claussen & Sahai (1992) discovered large amounts of molecular gas from observations of the $^{12}$CO(1–0) line. The line asymmetries in the CO line suggested that the galaxy is tidally disturbed. NGC 7552 is well known for its kpc-sized circumnuclear ring, which is the powering source of the nuclear starburst (Schinnerer et al. 1997; Brandl et al. 2012). The ring and the spiral arms starting at its circumference are largely obscured at optical wavelengths but are becoming more and more evident towards the near- and mid-IR. The ring is dominated by a group of UV-bright star clusters separated by tens of pc. There is also evidence for a bar within a bar but no signs of nuclear activity.

## 3. Observations and Data Reduction

We used the short-wavelength, medium-resolution far-UV mode (G130M) of COS to observe the program galaxies between May and October 2011. The HST program ID is 12173. A description of the COS spectrograph was given by Osterman et al. (2011) and Green et al. (2012). Details of the observations are in Table 2. This table lists the HST observation set identifiers, galaxy names, and observation epochs in columns 1, 2, and 3, respectively. Each galaxy was acquired with a near-UV imaging target acquisition through the primary science aperture (PSA). We specified the coordinates corresponding to the UV-brightest regions, as found from previous UV observations with HST or other telescopes. Since all galaxies are classified as nuclear starbursts, the location of the COS aperture is close to the galaxy center. We searched the Hubble Legacy Archive for existing UV images of the program galaxies. All four galaxies have a rich history of previous HST observations. In Figure 1 – 4 we show HST UV images of the four galaxies with the location of the PSA overplotted. The images for F08339+6517, NGC 3256, and NGC 6090 were obtained with the High Resolution Channel (HRC) of HST's Advanced Camera for Surveys (ACS) using the F220W filter. There is no corresponding archival image of NGC 7552; therefore we use a Wide Field and Planetary Camera 2 (WFPC2) image through the F336W filter. The wavelength of this filter is reasonably close to those of the COS far-UV channel to give a realistic rendering of the UV spatial morphology encountered by the COS PSA. F08339+6517 (Figure 1) has a compact morphology in the UV, with most of its UV light concentrated in the single nucleus (Margon et al. 1988). Therefore the COS PSA placement is straightforward. NGC 3256 (Figure 2) has a complex UV morphology. Whereas IR images show a bright compact starburst nucleus (Alonso et al. 2002), heavy dust extinction in the galaxy center generates several separate flux maxima in the UV. The COS spectra were obtained at the indicated location. It is important to realize that the UV fluxes encompassed by the PSA will only be a small fraction of what has been measured with large-aperture spectrographs. NGC 6090 (Figure 3) shows two distinct nuclei in the UV. We centered the COS aperture on the southeastern star-forming region, which has higher UV flux levels. Again, we point out that the COS fluxes will be lower than those observed by International Ultraviolet Observer (IUE) since the IUE aperture includes both nuclei.

NGC 7552 (Figure 4) does not have a single dominating star cluster in the UV. The COS spectrum was obtained near the nucleus inside the circumnuclear ring, which is bright in the IR but faint in the UV.

Following the successful acquisition and centering in the PSA, spectroscopic data were obtained at the default focal plane offset position (FP-POS = 3) in TIME-TAG mode for each galaxy. In order to provide continuous spectral coverage while minimizing the impact of micro-channel plate detector fixed-pattern noise, we observed each galaxy at two, three or four central wavelength positions (column 4 of Table 2). In column 5 of the table we provide the effective wavelength range in the galaxy restframe with this set-up, taking into account the redshift of each galaxy. The exposure times for each grating setting are given in column 6. The corresponding number of orbits spent on each galaxy are 1, 4, 3, and 2 for F08339+6517, NGC 3256, NGC 6090, and NGC 7552, respectively.

The individual exposures for each galaxy were retrieved from MAST and processed on-the-fly with version 2.15.4 of the CalCOS pipeline. CalCOS uses a series of modules which first correct the data for instrument effects such as detector noise, thermal drifts, geometric distortions, orbital Doppler shifts, count-rate non-linearity, and pixel-to-pixel variations in sensitivity. Then a standard wavelength scale is applied using the default onboard wavelength calibration. The final step is the extraction of flux-calibrated, combined spectra, with a heliocentric Doppler correction applied.

We further processed the retrieved data sets with IDL software developed by us and by the COS GTO team. Danforth et al. (2010) discussed an optimum exposure coaddition algorithm which we adopted as well. Their IDL routine reads in the individual x1d files and removes major flat-field artifacts from the flux vector, such as shadows from the ion repeller grid wires, and modifies the error and exposure time in these pixels. Then the exposures at the detector edges are de-weighted, and the individual exposures are cross-correlated and interpolated onto a common wavelength vector. The spectra were then resampled from the original dispersion of 0.00997 Å pix$^{-1}$ to 0.06 Å pix$^{-1}$, which corresponds to the nominal resolution. We checked the precision of the internal wavelength calibration using the geocoronal Lyman-α and O I lines at 1215.67 Å and 1302.17/1304.86/1306.03 Å, respectively. Offsets were found to be less than 1 pixel in all cases. Additional wavelength zero point offsets are introduced by centering errors and the non-uniform light distribution of the galaxies in the PSA. We determined these offsets by measuring the wavelengths of the Milky Way foreground absorption lines of Si III λ1206.50, S II λ1250.58, S II λ1253.81, Si II λ1260.42, and C II λ1334.53. The underlying assumption is that these lines are unshifted and not affected by Galactic high-velocity clouds (see Leitherer et al. 2011). We found typical wavelength offsets of ~0.2 Å and corrected the spectra accordingly.

We corrected the galaxy spectra for Milky Way foreground reddening and applied the heliocentric redshift corrections using the values in Table 1. The final step of our data reduction pipeline was a boxcar smoothing over five pixels to account for the degraded spectral resolution caused by the spatial extension of the targets in the PSA. We measured the widths of the Milky Way foreground lines, which are expected to be unresolved at the nominal COS G130M resolution of R ≈ 17,000. Typical widths were found to be ~0.5 ± 0.3 Å. We also generated rectified versions of the galaxy spectra, whose line-free regions were fitted with a multi-order polynomial and divided by the resulting continuum spectrum. In Figure 5 – 8 we show the rectified spectra, which were processed through all previously mentioned steps. We identified the major spectral features using the line list of Leitherer et al. (2011). The strongest emission lines are the geocoronal lines of N I λ1200, Lyman-α, and O I λ1304, as well as Lyman-α intrinsic to the galaxies. We identified the major Milky Way foreground contaminants at the bottom of each figure. The spectral features originating in the galaxies are identified at the top and will be discussed in more detail in the following sections.

## 4. Dust Reddening and Stellar Population

Although the star-formation histories of the four galaxies are not the primary focus of this study, we took advantage of the stellar lines in the UV spectra for a basic analysis of the stellar population properties. We applied the standard technique of UV spectral synthesis (e.g., Tremonti et al. 2001; Chandar et al. 2003; Wofford, Leitherer, & Chandar 2011) to constrain the most recent star-formation history traced by massive OB stars.

We adopted solar chemical composition and a standard Kroupa-type IMF (Kroupa 2008) between 0.1 and 120 $M_\odot$. This IMF approaches a Salpeter slope with power-law index $\alpha = 2.35$ (on linear mass intervals) at the high-mass end and has been found to be widely applicable in star-forming galaxies (Leitherer 2011). Since the COS PSA covers size scales of $10^2 - 10^3$ pc (see Table 1), it is reasonable to assume that the short-lived massive stars have reached an equilibrium between formation and death, and star-formations is quasi-continuous. In this case the UV spectrum becomes almost age-independent, and we adopt a characteristic age of 20 Myr, corresponding to the evolutionary time scale of a 10 $M_\odot$ star. At the other extreme, one might invoke a quasi-instantaneous star-formation history. In this case, the age of the burst would have to be less than a few Myr in order to be consistent with the presence of strong O-star features in the UV spectra, which would disappear at later epochs. Such a young age would be inconsistent with other indicators, most importantly with the presence of older red supergiants which have been found in all four galaxies. Accounting for the presence of both O stars and red supergiants requires starburst durations of at least ~10 Myr, in conflict with the single burst scenario.

We used the Starburst99 code for all modeling (Leitherer et al. 1999; Vázquez & Leitherer 2005; Leitherer & Chen 2009). We determined the intrinsic dust reddening by comparing the observed slope of the UV continuum to the theoretically expected value over the full wavelength range observed with the G130M grating. This technique takes advantage of the very weak dependence of the UV continuum slope on stellar properties so that the slope becomes a measure of the dust attenuation alone. The model spectra were calculated from the theoretical atmospheres of Smith, Norris, & Crowther (2002). We used the extragalactic reddening law of Calzetti et al. (2000), which is widely used for dust corrections of starburst spectral energy distributions. In Table 3 (column 2) we report the derived intrinsic dust reddening $E(B–V)_{int}$, which has been determined from the continuum slope. After dereddening the spectra with $E(B–V)_{tot}$, the sum of the Milky Way foreground and intrinsic reddening, and adopting $D$ of Table 1 we derived the monochromatic UV luminosities $L_{COS}$ (column 3 of Table 3). $L_{COS}$ refers to a wavelength of 1310 Å, which is a line-free region of the continuum. Since the spectra are quite flat, there is no strong dependence on wavelength. $L_{COS}$ is a direct proxy for the rate of massive stars forming in the current star-formation episode. By extrapolating to lower masses using the adopted IMF, we derived the total star-formation rates $SFR_{COS}$ listed in column 4. This is the star formation sampled by the COS aperture, which is a lower limit to the global rate seen in the UV because of the small size of the PSA. The global rates have been measured for the four galaxies with large-aperture spectrographs such as IUE or the Hopkins Ultraviolet Telescope (HUT). All four galaxies are classified as LIRGs ($10^{11} < L_{IR}/L_\odot < 10^{12}$) and have been found to follow the empirical relation between the far-IR luminosity and the UV reddening (Meurer et al. 1995; Meurer, Heckman, & Calzetti 1999; Leitherer et al. 2002). Therefore the dereddened UV luminosity $L_{UV}$ is equivalent to the stellar radiation absorbed in the UV and re-emitted in the far-IR so that $L_{UV} \approx L_{IR}$. In other words, the observed reddening corrected UV luminosity accounts for the entire star formation – there is no significant starburst totally hidden by dust. The relationship between the radiation absorbed in the UV and re-emitted in the IR strongly depends on the bolometric luminosity of the star-forming galaxy (Goldader et al. 2002; Howell et al. 2010). At luminosities below $10^{10} – 10^{11}$ L$_\odot$, the reddening corrected UV luminosity serves as a good proxy for the star-formation rate, whereas at luminosities above $10^{12}$ L$_\odot$ star formation is so highly attenuated that the UV does not trace star formation at all. LIRGs have luminosities that place them between these two extremes: The contribution of the UV to the measured star-formation rate is on average 4%, with a large variation (Howell et al.) suggesting most star formation is deeply embedded but some fraction is available to direct observations. The four galaxies in our study have properties closer to those of less luminous galaxies as implied by our ability to recover the absorbed UV radiation from the IR.

We compared the observed COS fluxes to those observed with IUE in order to assess the fraction of star formation sampled by COS. The assumption is made that the IUE aperture samples the entire recent star formation of the four galaxies. The IUE spectra of the four galaxies were retrieved from MAST for this purpose. In column 5 of Table 3 we give $SFR_{IUE}$, the star-formation rates derived from the 1310 Å fluxes through the IUE aperture. (The IUE aperture has dimensions 10″ by 20″.) Not surprisingly, the ratios of $SFR_{COS}/SFR_{IUE}$ are smallest for NGC 3256 and NGC 7552, whose UV morphologies are determined by dust and a ring-geometry. In contrast, the COS PSA samples close to ~40% of the IUE fluxes in F08339+6517 and NGC 6090 which have rather centrally concentrated morphologies.

Using the derived stellar population properties we computed synthetic line spectra for comparison with the stellar features in the COS spectra. With the previously mentioned parameters fixed, there are no additional tunable inputs, and the line spectrum becomes an important consistency test. In Figure 9 – 12 we provide the comparison between the observed, dereddened, and theoretical spectra of the four galaxies. The observed spectra at the bottom of each figure correspond to those reproduced in Figure 5 – 8, except for not being rectified and corrected for foreground reddening. The theoretical spectra were generated with an empirical library of Galactic stars which contain weak interstellar absorption lines but do otherwise not display any signatures of galactic winds. Therefore one can expect agreement between the observed and synthetic spectra only for the stellar lines. The weak, blended continuum features are in good agreement (e.g., the broad O IV blend at ~1340 Å). Stellar-wind lines are more difficult to assess. The strongest, most commonly used stellar-wind line of C IV λ1550 is outside the covered wavelength range. N V λ1240 can provide constraints instead. It is a P Cygni profile whose absorption part is strongly affected by Galactic interstellar Lyman-α absorption in the library spectra. Limiting the comparison to the emission part of the line suggests a very good match between the models and the observations. Since N V (like other stellar-wind lines) is quite sensitive to IMF variations (Leitherer 2011), the agreement supports a Kroupa IMF for massive stars in the four galaxies. $Si^{3+}$ has a lower ionization energy of 33 eV, compared to 77 eV for $N^{4+}$. As a result, N V has less interstellar contribution to the wind line, whereas the Si IV λ1400 doublet is dominated by the non-stellar component. The broad stellar emission component of Si IV λ1400 is again well reproduced by the models. The absorption in the synthetic spectra is due to a combination of unshifted Galactic interstellar gas and weak, broad (up to ~2000 km $s^{-1}$) stellar-wind absorption. The Si IV absorption lines observed in the four galaxies are strikingly different from the models. They are stronger and broader than the synthetic lines, and have a higher bulk (not necessarily maximum) velocity. The same behavior can be seen in the other interstellar lines, such as Si III λ1206, Si II λ1260, or C II λ1334. As we will discuss in the following section, these lines are the signatures of strong galactic-scale outflows driven by the powerful starbursts.

To summarize, the stellar lines indicate galaxy properties similar to those derived from UV spectra of less luminous, less dusty, and more metal-poor galaxies. However, this does not hold for non-stellar lines, which are much stronger relative to stellar lines in the IR-luminous sample when compared with less luminous galaxies. Stellar-wind lines in less luminous galaxies are generally stronger than interstellar and circumgalactic absorption lines, whereas in our IR-luminous sample the opposite is true: the UV spectrum is dominated by strong non-stellar lines, and stellar lines are far less conspicuous. The stellar lines would be difficult to detect in UV spectra at lower S/N and spectral resolution. This situation is reminiscent of observations at high *z*, where the spectral features in the restframe UV are mostly non-stellar (e.g., Steidel et al. 2010).

## 5. Lyman-α

Lyman-α is the strongest intrinsic emission line in all four galaxies. The redshift of the galaxies is sufficiently high to separate the intrinsic Lyman-α from geo-coronal emission and Galactic interstellar absorption. A comparison of the profiles is in Figure 13 where we display zoomed sections of the normalized spectra. Lyman-α has a dominant, redshifted emission component, with blueshifted absorption extending over several hundred km s$^{-1}$. The Lyman-α profiles of F08339+6517 and NGC 6090 were discussed before by Kunth et al. (1998) and González Delgado et al. (1998) who emphasized the importance of the velocity structure of the ISM over metallicity and dust content for the escape probability of Lyman-α photons. Dust is expected to play at least some role due to the resonant nature of Lyman-α (Finkelstein et al. 2011). The photons scatter in the neutral ISM of their host galaxies, thereby significantly increasing their path lengths and their sensitivity to absorption by interstellar dust. In the absence of other effects, this implies a significant reduction of Lyman-α photons in dust-rich galaxies (Charlot & Fall 1993). Our data support the presence of additional escape mechanisms, such as the ISM velocity structure. We measured the equivalent widths (*EW*) of Lyman-α in the normalized spectra between 1214 and 1219 Å. The boundaries are somewhat arbitrary since the profile is a composite of the strong emission, underlying absorption, and the red wing of the Galactic interstellar damped Lyman-α. Our choice of the boundaries essentially measures the emission component. We found $EW$ = 5.6, 13.3, 3.2, and 2.1 Å for F08339+6517, NGC 3256, NGC 6090, and NGC 7552, respectively. This can be compared with the theoretically predicted value of $EW$ = 101 Å for a 20 Myr old continuous starburst, identical dust attenuation for stars and gas, no H I scattering of Lyman-α photons, and Case B recombination (i.e, all Lyman lines are optically thick and Lyman-α/Hα = 8.1). Note that this prediction is very insensitive to the adopted age for ages above ~10 Myr, as can be seen in Figure 15 of Verhamme et al. (2008). The observed values reach 2 – 13% of the model numbers, indicating a substantial destruction rate of Lyman-α photons.

We did not account for any underlying stellar component. Observational data for stellar Lyman-α in hot stars are scarce because even the closest OB stars are affected by significant interstellar H I absorption which dominates any stellar Lyman-α (Diplas & Savage 1994a,b). As a result, empirical stellar Lyman-α profiles exist only for B stars, but not for O stars (e.g., Savage & Panek 1974; Vader, Pottasch, & Bohlin 1977). These observations suggest Lyman-α $EW$'s of tens of Å in absorption in B stars, with strongly decreasing values for hotter stars. Are B stars expected to be the main stellar contributor to the underlying Lyman-α in our galaxies? The spectral morphology suggests otherwise. The four galaxy spectra show conspicuous, broad N V λ1240, which is strong in individual O stars but absent in the spectra of stars later than B0 (Walborn, Bohlin, & Panek 1985; Walborn, Parker, & Nichols 1995). Furthermore, B stars in young populations can be identified via several *excited* transitions, such as, e.g., Si II λ1265 (de Mello, Leitherer, & Heckman 2000), none of which are detected in our spectra. It is therefore unlikely for B stars to make a major contribution to stellar Lyman-α. Rather, O stars are the main contributor, and we need to turn to models to evaluate their Lyman-α strength. Klein & Castor (1978) computed a small set of unblanketed non-LTE atmospheres for O stars with temperatures between 30,000 and 50,000 K and found Lyman-α *EW*'s between 0.2 Å in emission and 1.2 Å in absorption. In order to verify these values with more modern, blanketed model atmospheres, we measured the Lyman-α line in the models of Bouret et al. (2012). The spectral region around Lyman-α in their models is reproduced in Figure 14. The seven individual stars are HD 14947 (O4.5 I), HD 15570 (O4 I), HD 66811 (O4 I), HD 163958 (O6.5 I), HD 190429A (O4 I), HD 192629 (O7.5 I), and HD 210839 (O6 I). The individual profiles show P Cygni structures with relatively weak absorption and emission. The net *EW* is close to zero because emission and absorption essentially cancel each other. The average spectrum of the seven models (corresponding to a representative early- to mid-O star) has *EW* = 0.24 Å in absorption, which is consistent with the earlier results of Klein & Castor. A full spectral synthesis of Lyman-α will be a future effort (Peña-Guerrero & Leitherer, in preparation) but our simple estimates suggest that any correction for underlying stellar Lyman-α will be negligible.

Although our small sample size precludes general conclusions, we note the absence of any correlation with reddening. NGC 3256 and NGC 7552 have the largest $E(B-V)_{int}$ of 0.42 and 0.53, respectively, but at the same time have the largest and the smallest *EW*, respectively. Shapley et al. (2003) carried out a systematic study of the rest-frame UV spectroscopic properties of Lyman-break galaxies. They found a median Lyman-α *EW* of ~0 Å in their sample of ~1000 galaxies. Our galaxies fit well into the range of *EW* values observed for their sample. While both samples have similar luminosities and star-formation rates, our IR-luminous sample has substantially higher dust content. The Shapley sample has a mean *E(B−V)* of 0.13, much less than for our sample (see Table 3). Therefore it is not unexpected that our sample does not follow a trend of *EW* with *E(B−V)* as found by Shapley et al. in their Lyman-break sample (their Fig. 10).

In order to gain further insight into the physics governing the observed Lyman-α profiles, we calculated a set of synthetic Lyman-α lines using the models developed by Schaerer et al. (2011). These models are an improved version of the Monte Carlo radiation transfer code Verhamme, Schaerer, & Maselli (2006), which accounts for the detailed physics of the Lyman-α line and UV continuum transfer, dust scattering, and dust absorption for arbitrary 3-D geometries and velocity fields. Here we assumed a spherical shell expanding towards the observer at constant velocity $v_{exp}$ and with an intrinsic velocity distribution described by the Doppler parameter *b*. The shell has a total neutral hydrogen column density *N(H)* and a dust optical depth $\tau_a$. We varied these four parameters over the entire parameter space described in Schaerer et al., i.e., $0 < v_{exp} <$ 700 km s$^{-1}$, $10^{16} < N(H) < 10^{22}$ cm$^{-2}$, $0 < \tau_a < 4$, and $10 < b < 160$ km s$^{-1}$. The intrinsic Lyman-α equivalent width is varied between 0 and 300 Å, and the line is assumed to be a Gaussian with a full width at half maximum of 100 km s$^{-1}$. Using our automatic fitting engine developed by Matthew Hayes, we have then determined the best-fitting profiles for the four galaxies. Previous applications of our fit procedure to high-*z* galaxies are discussed in Dessauges-Zavadsky et al. (2010), Vanzella et al. (2010), and Lidman et al. (2012). The best-fit profiles for F08339+6517, NGC 3256, NGC 6090, and NGC 7552 are reproduced in Figure 15, Figure 16, Figure 17, and Figure 18, respectively. In each figure we list the parameters used for the synthetic profiles. While not perfect, the fits are reasonably good to give some confidence in the modeling. Note that at wavelengths shortward of the line center the observed continuum can be affected by the red wing of the blueshifted Galactic damped Lyman-α absorption. This can be readily seen in Figure 5 – Figure 8, which display the wavelength region around Lyman-α over a wider range. Galactic and geocoronal Lyman-α clearly are a concern for NGC 3256 and NGC 7552, the two galaxies with the lowest redshift. Furthermore, intrinsic Si III λ1206 as well as blueshifted Galactic S II λ1250 (in NGC 6090; see Figure 7) affected the observed continuum location. Therefore the mismatch between the observed and theoretical profiles is expected and not indication of a failure of the models. The best fits to the Lyman-α profiles of the four galaxies have log $N(H)$ between 19.3 and 20.2, $v_{exp}$ between 50 and 150 km s$^{-1}$, $b$ between 10 and 40 km s$^{-1}$, and $\tau_a$ between 0 and 2. These are values are very similar to those derived in Lyman-break galaxies at $z \approx 3$ (Verhamme et al. 2008). However, the analogy between our sample and Lyman-break galaxies should not be pushed too far. In the case of distant galaxies, the spectrograph aperture typically encompasses the entire Lyman-α emission of the galaxy, whereas here we only sample the central region. Östlin et al. (2009) discussed HST/ACS Lyman-α images of star-forming galaxies, including F08339+6517 and NGC 6090. Both galaxies show extended Lyman-α emission from regions clearly outside the COS aperture, whereas our schematic model assumes a point-like emission source surrounded by a spherical shell. Therefore the derived profile-fit parameters should be taken with care.

## 6. Kinematics of the interstellar absorption lines

The most conspicuous absorption features in all four galaxies are the low-ionization lines of Si III λ1206, Si II λ1260, and C II λ1334. Being resonance transitions, these lines could be formed in stellar winds and/or the ISM. If the lines originated in winds, the mass losing stars would be of spectral type B (see the extensive discussion in de Mello, Leitherer, & Heckman 2000). Si II becomes strongest in late-B supergiants, whereas C II and Si III peak in early-B supergiants. Such stars are not expected to contribute strongly to the UV luminosity in the populations of the four galaxies. O stars dominate the light as suggested by our population synthesis models and supported empirically by a straightforward test pointed out by de Mello et al.: individual B stars show both the resonance line of Si II λ1260 and the excited transition of Si II λ1265.04. The absence of the latter line indicates a population in which B stars are not significant contributors to the resonance line, and it is predominantly interstellar.

Inspection of the line profiles of Si III, Si II, and C II suggests a complex structure with several components present. We used the corresponding Galactic foreground lines as fiducials to verify that the observed profiles are not artifacts caused by the morphology of the star-forming regions encompassed by the COS aperture. The observed line profiles result from the intrinsic line-spread function and the spatial profile of the source inside the COS aperture. Therefore the observed complex profile structure could be caused by both velocity structure of the source or the source morphology in the aperture. The Galactic foreground absorption lines are consistent with a single Gaussian with FWHM of 0.5 Å, and no indication of additional components. This suggests velocity structure as the origin of the multiple components seen in the Si III, Si II, and C II profiles of the four galaxies.

Among the three lines C II λ1334 is the strongest in the four galaxies. The overall profiles are blueshifted and composed of several discrete components with velocities of up to almost ~900 km s$^{-1}$. This is illustrated in Figure 19, Figure 20, Figure 21, and Figure 22, where we plot C II, Si III λ1206, and Si II λ1260 in velocity space with the detected components highlighted. The velocities refer to the wavelengths of the flux centroids in C II λ1334 measured in each galaxy. The reality of the detected features can be verified by comparison with the other two low-ionization lines Si III λ1206 and Si II λ1260, which are also included in these figures and to which we compared the C II velocity measurements. We chose a conservative approach and dismissed structures whose origin could not clearly be attributed to a high-velocity component. For instance, a strong absorption at 1324 Å is more likely due to N III λ1324.35 rather than blueshifted C II. However, given the redshift of the four galaxies, we can confidently rule out any confusion with Galactic high-velocity clouds. Note that the strong emission at the wavelength of Si III λ1206 in NGC 3256 (Figure 20) is geocoronal Lyman-α.

The velocity of the highest velocity material does not necessarily represent the average velocity of the bulk material. In order to determine a more conservative estimate of the flow velocity, we fitted single Gaussians to the total Si III, Si II, and C II profiles with central wavelength as a free parameter. We obtained velocities of –438, –461, –232, and –353 km s$^{-1}$ for F08339+6517, NGC 3256, NGC 6090, and NGC 7552, respectively.

Next we turn to higher-ionization lines. Si IV λλ1393,1402 is a mixture of stellar-wind and interstellar components. The contribution from winds is obvious from the emission part of the P Cygni profiles of Si IV in all four galaxies. Moreover, there is a smooth, blue wing extending up to ~2000 km s$^{-1}$ in Si IV. This can easily be seen in the comparison between the observed and synthetic spectra reproduced in Figure 9 through Figure 12. The blue wing is a tell-tale sign of *stellar* winds and should not be confused with *galactic-scale* outflows. Figure 23 through Figure 26 reproduce the Si IV lines of the four galaxies in the same way as the low-ionization lines in the previous figures. We included the velocity measurements as obtained from the C II lines to help identify any absorption components at similar velocities. The velocities of the individual components agree remarkably well with the velocities observed in the low-ionization lines. This gives confidence in an interpretation as due to outflowing gas.

N V λλ1238,43 has an ionization potential of 77 eV. Together with O VI λλ1031,37, C IV λλ1548,50 (both are outside the wavelength range covered here) and Si IV λλ1393,1402, N V is the strongest stellar-wind feature in the UV spectra of star-forming galaxies. Any interstellar feature would be hard to detect in the presence of strong N V wind absorption, given the achievable spectral resolution and S/N of HST's (and other) UV spectrographs. The N V lines in Figure 23, Figure 24, Figure 25, and Figure 26 confirm this expectation. There is no convincing evidence of absorption features at the same velocity in both N V components. Note that the strong absorption in N V λ1238 in F08339+6517is blueshifted Galactic Si II λ1260, and the feature in N V λ1242 at −126 km s$^{-1}$ in NGC 3256 is blueshifted Galactic Si II λ1253.

The most straightforward interpretation of the observed velocity structure is expanding warm (~$10^4$ K) gas driven by the current burst of star formation. The individual components indicate density inhomogeneities of the outflow, as expected for interstellar material entrained in a galactic-scale wind. Using $v = 900$ km s$^{-1}$ for the maximum outflow velocity and a maximum distance of $d = 10$ kpc (corresponding to the typical sizes of galactic winds seen in starburst galaxies [Heckman et al. 2000]), we find a time scale of $t = d/v \approx 10$ Myr, which corresponds to the typical durations of starbursts (Gallagher & Smith 2005). Does the material escape from the gravitational well of the galaxies? We first compare the outflow velocities to the observed rotation velocities of the four galaxies. The relevant information is in Table 4. In column 2 of this table we list $W_{20}$, the full width at 20% peak intensity of the H I profile as published by the references in column 3. $W_{20}$ was converted to a rotation amplitude $W_R$ (column 4) using equation (12) of Tully & Fouque (1985) with their default parameters. The inclinations $i$ of the four galaxies are small. We estimated $i$ by comparing the expected ratio $b/a = 0.2$ of the minor over the major axis with the values found on NED (column 5). Column 6 gives the rotation velocities $W_{rot}$, derived from half the rotation amplitude $W_R$ and corrected for inclination. F08339+6517 has a peculiar H I morphology with evidence of H I being stripped by interaction with a nearby companion galaxy (Cannon et al. 2004). Therefore the H I line width may not indicate rotation. In fact, applying our prescription to this galaxy would lead to an unreasonably high rotation velocity of 667 km s$^{-1}$. Therefore we discarded this value and assigned the average rotation velocity of the other three galaxies to F08339+6517.

The observed maximum outflow velocities clearly exceed $v_{esc} = \sqrt{2}\ W_{rot}$, the minimum requirement for escape if the galaxies were point masses. A more realistic approximation is an isothermal gravitational potential that extends to a maximum halo radius (Binney & Tremaine 1987). In this case, $v_{esc}$ is related to the rotation velocity as $v_{esc} \approx 2.75\ W_{rot}$ (Heckman et al. 2000). The halo size has only a mild effect on the scaling factor between $v_{esc}$ and $W_{rot}$ which will always be between 2.5 and 3. Interestingly, the maximum outflow velocities of the four galaxies are quite similar to the escape velocities. Therefore it is not clear whether the material will leave the galaxy potential well or will fall back to the galaxy disk. Either way, the removal of the gas will have profound implications for the star formation process.

We can quantify the outflow rates associated with the gas observed in the UV spectra using the formulation of Rupke et al. (2005b) and Weiner et al. (2009). Assuming a thin shell at a characteristic distance $r$ from the central starburst having a column density $N(H)$ and velocity $v$, the outflow rate can be written as

$$\dot{M} = \Omega \mu m_p N(H) r v, \tag{3}$$

where $\mu$ and $m_p$ are the mean molecular weight and the proton mass, respectively. Here we adopt $\mu = 1.4$. The solid angle of the wind as seen from the central starburst $\Omega$ can be expressed in terms of an angular covering fraction $C_\Omega$ and a clump covering fraction $C_f$:

$$\Omega = 4\pi C_\Omega C_f. \tag{4}$$

$C_f$ accounts for the existence of individual discrete structures along the sightline, as suggested by the multiple components present in the UV absorption lines. Theoretically, such structure can be understood in terms of shells fragmented by Rayleigh–Taylor instability as they accelerate in the stratified ISM of the galactic disk (Fujita et al. 2009).

We can express the outflow rate more conveniently following Weiner et al.:

$$\dot{M} = 22 \text{ M}_\odot \text{yr}^{-1} C_\Omega C_f \frac{N(H)}{10^{20} \text{ cm}^{-2}} \frac{r}{5 \text{ kpc}} \frac{v}{300 \text{ km s}^{-1}}. \tag{5}$$

For an estimate of $\dot{M}$ of the four galaxies we adopted a characteristic distance of 5 kpc and a velocity of 400 km s$^{-1}$ (corresponding to the average of the bulk outflow velocity measured for the four galaxies). Rupke et al. (2005b) studied outflows in a sample of IR-luminous starburst galaxies in the optical and determined an average covering fraction $C_\Omega C_f = 0.15$, which we assumed as well. Ideally, one would like to determine hydrogen column densities for each of the four galaxies individually. For the reasons discussed before, the *N(H)* values suggested by the Lyman-α modeling are not reliable. On the other hand, the numerous ISM absorption lines observed in our UV spectra are not useful, either: the strong lines are heavily saturated, and all unsaturated lines have too low signal-to-noise for meaningful measurements. Therefore we used the average Galactic relation between *N(H)* and *E(B−V)* for estimates of the hydrogen column densities in the four galaxies:

$$N(H) = 4.9 \times 10^{21} \, E(B-V) \text{ cm}^{-2} \tag{6}$$

(Diplas & Savage 1994b). The underlying assumption for this relation to apply is that the dust reddening is dominated by the dust in the outflow, as opposed to dust in the galactic disk. Large amounts of dust in galaxy outflows have indeed been confirmed (Heckman et al. 2000), and their cosmological significance as the source of dust observed in Mg II absorbers has been demonstrated by Ménard & Fukugita (2012). A further assumption is that all hydrogen is in the form of H I and that H II and $H_2$ can be neglected. Using the measured *intrinsic* (i.e., the values in Table 3) reddening, we found rather similar *N(H)* values, with an average of $N(H) = (1.9 \pm 0.6) \times 10^{21}$ cm$^{-2}$. The value is typical for observed hydrogen column densities in star-forming galaxies (e.g., the rogue gallery of Hibbard et al. 2001) and agrees with direct observations of NGC 3256 (Maybhate et al. 2007), the only galaxy for which we found data in the literature.

Equation (5) suggests mass outflow rates of ~100 M$_\odot$ yr$^{-1}$ for the four galaxies. These rates can be compared to the total star-formation rates indicated by the far-IR luminosities. $L_{IR}$ is related to the star-formation rate via the relation

$$\log SFR = \log L_{IR} - 9.9, \qquad (7)$$

where $SFR$ is in M$_\odot$ yr$^{-1}$ and $L_{IR}$ is in L$_\odot$. As before, a Kroupa IMF has been adopted. Equation (7) follows from Starburst99 and is almost identical to the conversion relation given by Kennicutt & Evans (2012; their equation [12] and their Table 1). log $L_{IR}$ of the four galaxies ranges between 11.1 and 11.6 (Table 1) so that the corresponding star-formation rates are between 16 and 55 M$_\odot$ yr$^{-1}$, with a mean value of 29 M$_\odot$ yr$^{-1}$. The derived outflow rates are comparable to, or even higher than the star-formation rates. Galactic winds with outflow rates comparable to or exceeding the star-formation rates of starburst galaxies are commonly observed. Such winds are an important feedback mechanism by depleting the available gas reservoir and quenching star formation.

## 7. Conclusions

We present UV observations with COS of four IR-luminous galaxies. Despite their IR luminosities in excess of $10^{11}$L$_\odot$, the galaxy nuclei are sufficiently UV-bright to permit medium-resolution UV spectroscopy. This seemingly paradoxically result is the consequence of their porous, inhomogeneous ISM, leading to significant (non-ionizing) UV photon escape. All four galaxies were previously shown to follow the mean relation between $L_{IR}$ and UV spectral slope β. In general, this relation is not a reliable predictor of star-formation rates in IR-luminous galaxies as demonstrated in the GOALS survey by Howell et al. (2010). However, the galaxies in the present sample were consciously selected on the basis of high UV flux and are therefore biased in favor of following the $L_{IR}$ vs. β relation. While many IR-luminous galaxies are still detectable at UV wavelengths, this situation rapidly changes towards even higher IR luminosity. Goldader et al. (2002) found that the nuclei of ultra-luminous IR galaxies ($L_{IR} > 10^{12}$ L$_\odot$) are heavily dust-obscured and that even after correction for dust reddening, their reprocessed UV light is insufficient by large factors to account for the far-IR emission.

Most previous spectroscopic UV studies of star-forming and starburst galaxies concentrated on less luminous and (by virtue of the luminosity/metallicity relation) more metal-poor galaxies (e.g., Chandar et al. 2005; Leitherer et al. 2011). While star-forming dwarf galaxies deserve attention in their own right, more luminous galaxies are of particular interest because they more closely resemble their high-$z$ counterparts, such as the Lyman-break category. The most obvious difference between the IR-luminous sample studies here and Lyman-break galaxies is the dust content, which needs to be kept in mind when making comparisons.

We used the COS spectra to determine the stellar-population properties in the central kpc of the four galaxies. The internal dust reddening for a Calzetti-type attenuation law ranges between $E(B-V)_{int} = 0.25$ and 0.53, and the reddening-corrected star-formation rates within the COS aperture are between 1.1 and 26 M$_\odot$ yr$^{-1}$. The values are dependent on the projected COS aperture sizes and becomes quite similar for all four galaxies after normalization to identical surface areas. The total UV-derived star-formation rates through the IUE aperture are between 29 and 70 $M_\odot$ yr$^{-1}$. The observed stellar-wind lines are in good agreement with profiles predicted by population synthesis models and indicate that the high-mass stars follow a normal Kroupa-type IMF.

All four galaxies have Lyman-α in emission with equivalent widths ranging between 2 and 13 Å. These values are among the highest found in local star-forming galaxies (Wofford, Leitherer, & Salzer 2013), indicating significant escape of Lyman-α radiation. Nevertheless, even after correcting for any underlying stellar absorption (which will be small because OB stars dominate the continuum), the nebular Lyman-α is significantly weaker than the theoretically expected recombination value of ~100 Å. This makes Lyman-α a poor star-formation indicator in these galaxies.

The COS spectra convincingly show the presence of outflowing gas in all four galaxies. Blue-shifted absorption is observed in C II, Si II, and Si III at bulk velocities of ~400 km s$^{-1}$. There is no significant detection of a bulk outflow in lines associated with higher ionization energies, such as Si IV and N V. However, these lines have strong stellar-wind profiles, which may mask the signatures of outflowing gas. The line profiles of C II, Si II, and Si III have complex structures and split into at least three components whose maximum velocities reach up to ~900 km s$^{-1}$. The inhomogeneous ISM structure suggested by the absorption lines is the likely reason for the escape of Lyman-α photons, whose escape probability can increase in the presence of non-uniformities and velocity structures. On the other hand, Lyman continuum radiation does not follow such a simple pattern. F08339+6517 and NGC 6090 have been observed below 912 Å with the Hopkins Ultraviolet Telescope (HUT) with no significant detection (Leitherer et al. 1995).

The outflow velocities are much higher than typically observed in local starburst galaxies. Part of the reason for the high velocities is the inclination of the galaxies. All four are observed almost face-on, which maximizes the projected velocity of outflows along the galaxy minor axis. More importantly, the objects are among the most luminous local starburst galaxies with available UV spectroscopy. Our result suggests higher outflow velocities at higher luminosity. Despite having such outflow velocities, it is unlikely that the material will escape from the gravitational potential of the galaxies. We estimate that the maximum outflow velocities may approach but do not significantly exceed the escape velocities. Therefore the outflows are unlikely to be a source of enrichment for the IGM but will act as a reservoir of metals for the galaxy halos, similar to what has been suggested before by Tumlinson et al. (2011).

We obtained quantitative estimates of the outflow rates using the formulation of Rupke et al. (2005b) and Weiner et al. (2009). The derived rates of ~100 $M_\odot$ yr$^{-1}$ are comparable to or somewhat above the total IR derived star-formation rates. Therefore the outflows are an important mechanism for draining the gas reservoir and regulating star formation.

*Acknowledgments.* We are grateful to Jean-Claude Bouret for providing electronic versions of his CMFGEN O-star spectra. Maria Peña-Guerrero kindly provided assistance with the theoretical stellar Lyman-α profiles. Helpful comments from an anonymous referee significantly improved the paper. Support for this work has been provided by NASA through grant number GO-12173.01 from the Space Telescope Science Institute,

**Figures**

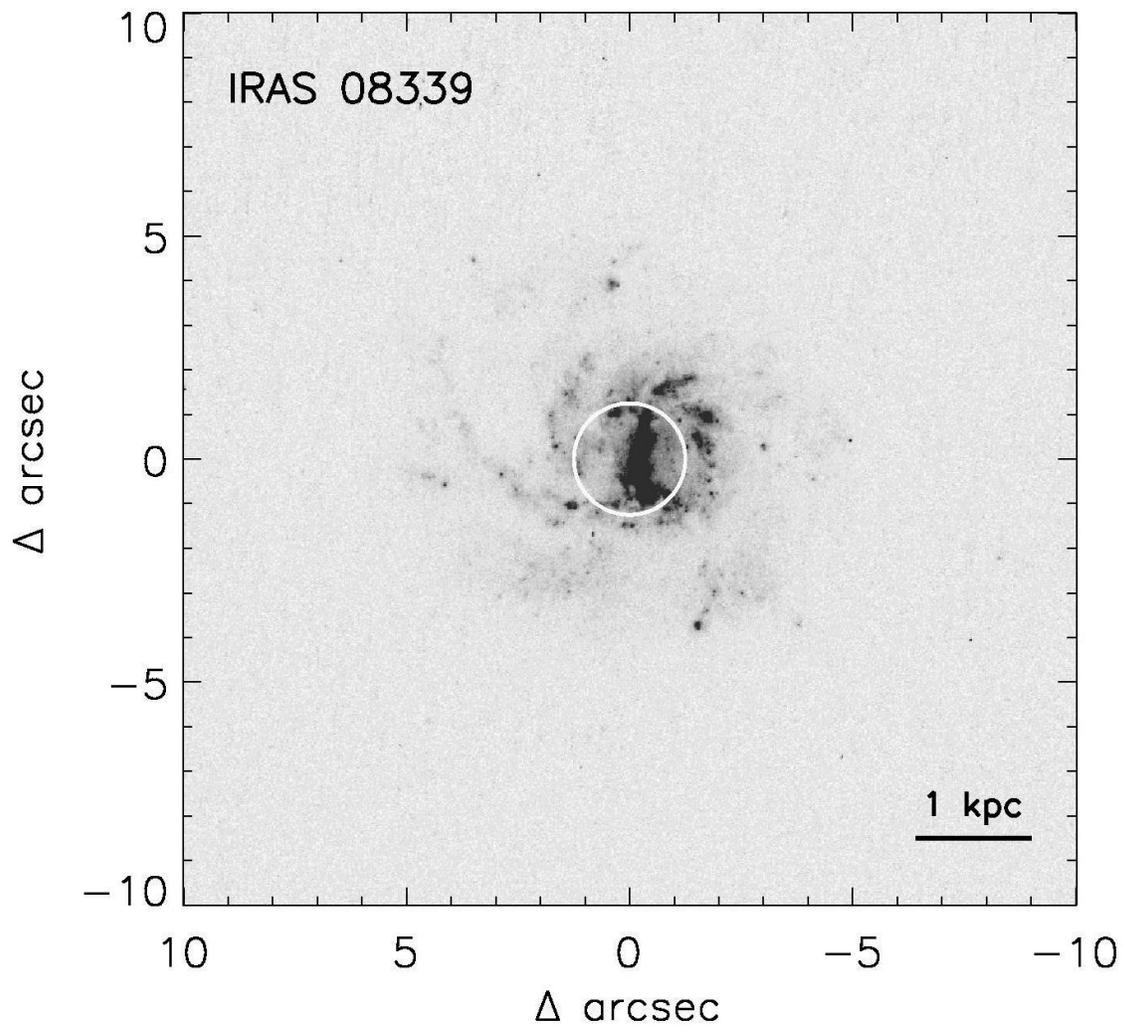

Figure 1. – HST ACS HRC image of F08339+6517 taken with the F220W filter. The image center is at RA = 08h38m23.15s and Dec = +65°07′15.4″. The location of the COS aperture is indicated by the circle. The linear scale is indicated by the bar in the lower right.

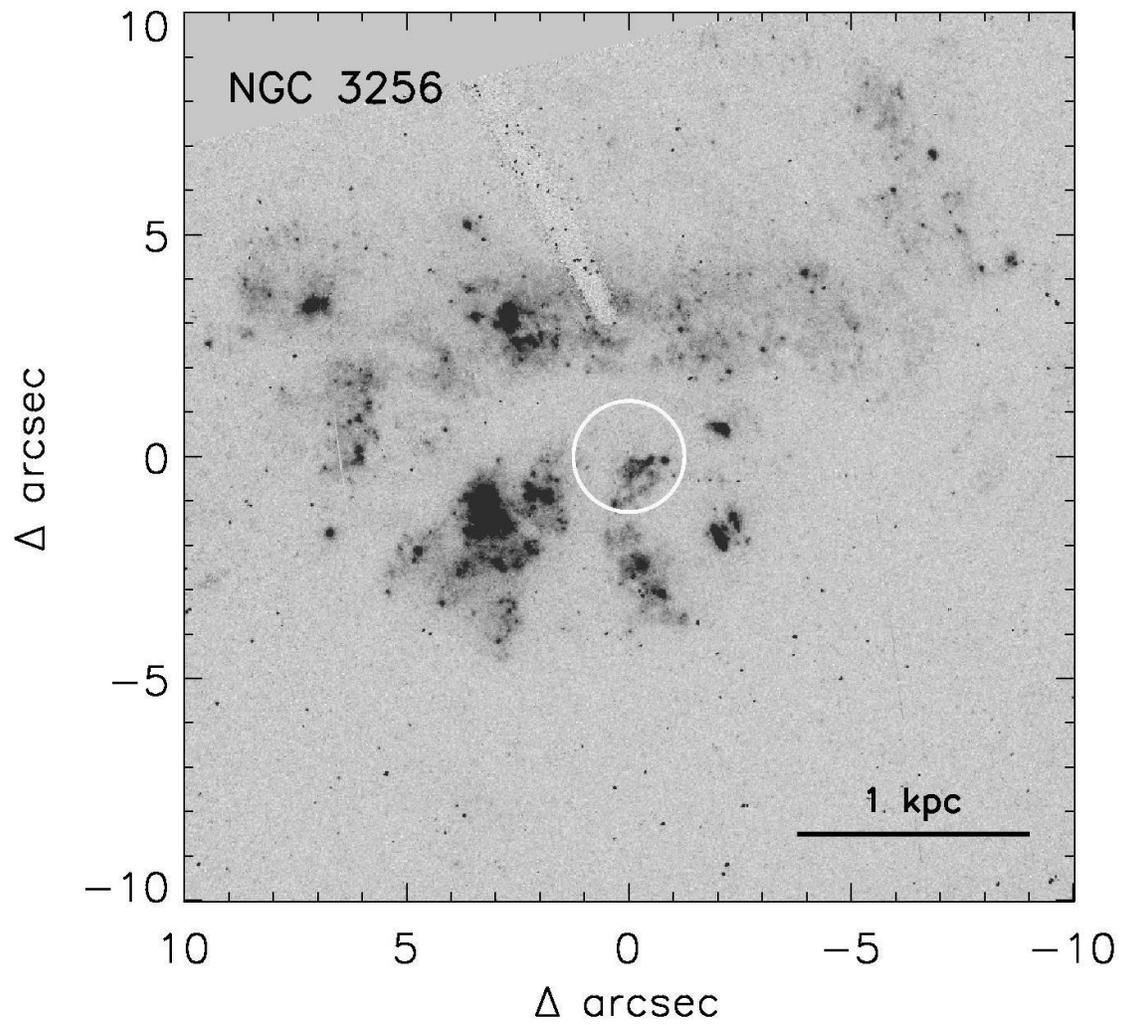

Figure 2. − HST ACS HRC image of NGC 3256 taken with the F220W filter. The image center is at RA = 10h27m51.34s and Dec = −43°54′12.4″. The location of the COS aperture is indicated by the circle. The linear scale is indicated by the bar in the lower right.

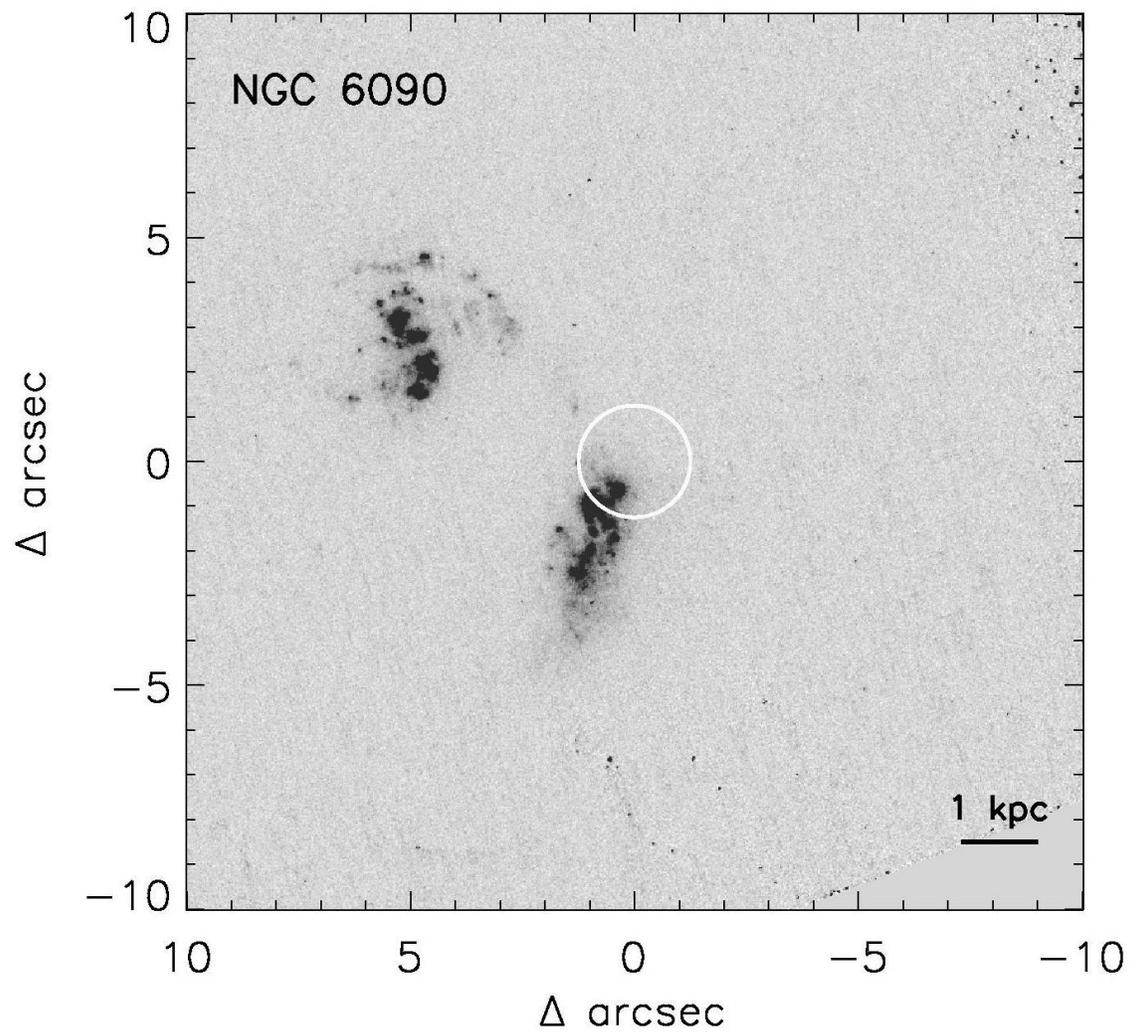

Figure 3. – HST ACS HRC image of NGC 6090 taken with the F220W filter. The image center is at RA = 16h11m40.20s and Dec = +52°27′23.8″. The location of the COS aperture is indicated by the circle. The linear scale is indicated by the bar in the lower right.

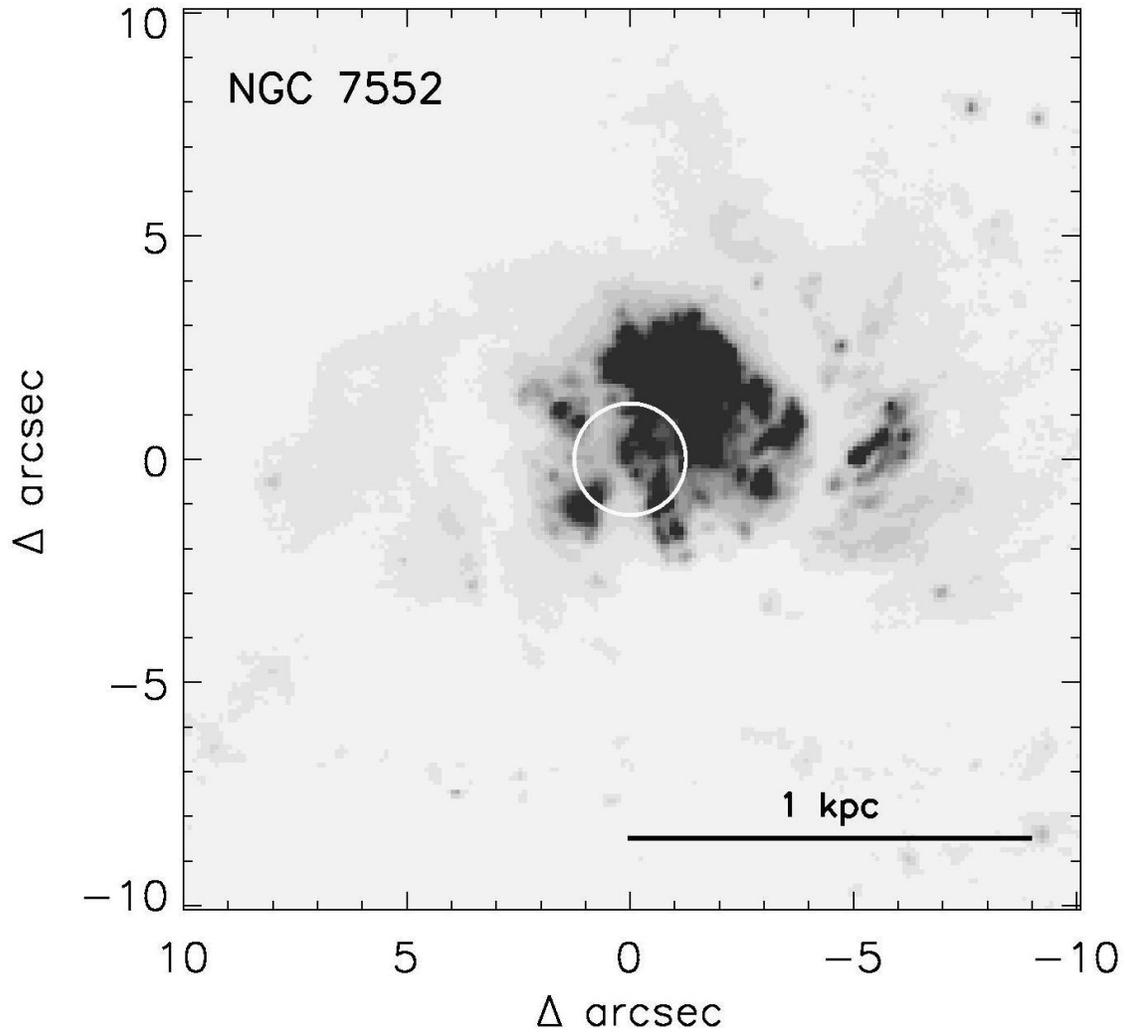

Figure 4. – HST WFPC2 image of NGC 7552 taken with the F336W filter. The image center is at RA = 23h16m10.79s and Dec = −42°35′05.5″. The location of the COS aperture is indicated by the circle. The linear scale is indicated by the bar in the lower right.

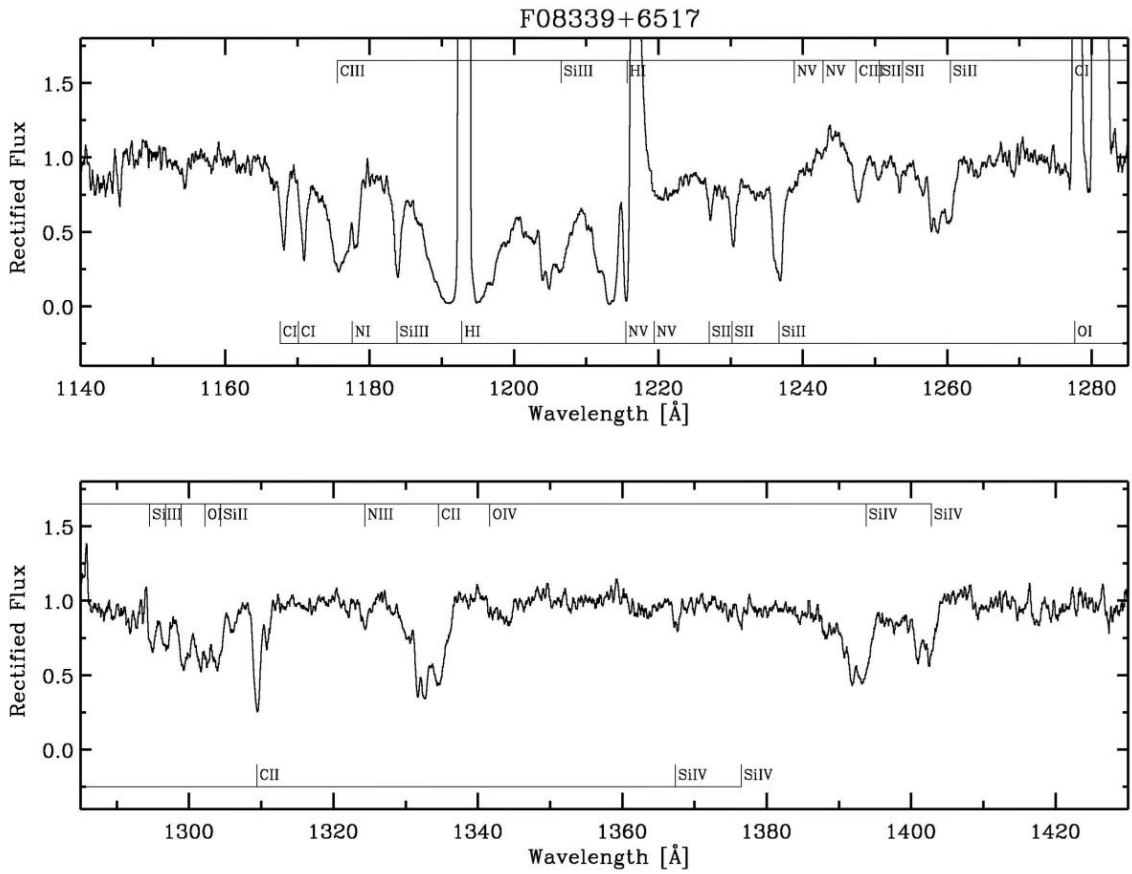

Figure 5. − Rectified COS spectrum of F08339+6517. The wavelength scale is in the restframe of the galaxy. Milky Way foreground absorptions are identified below the spectrum; major features intrinsic to the galaxy are identified above.

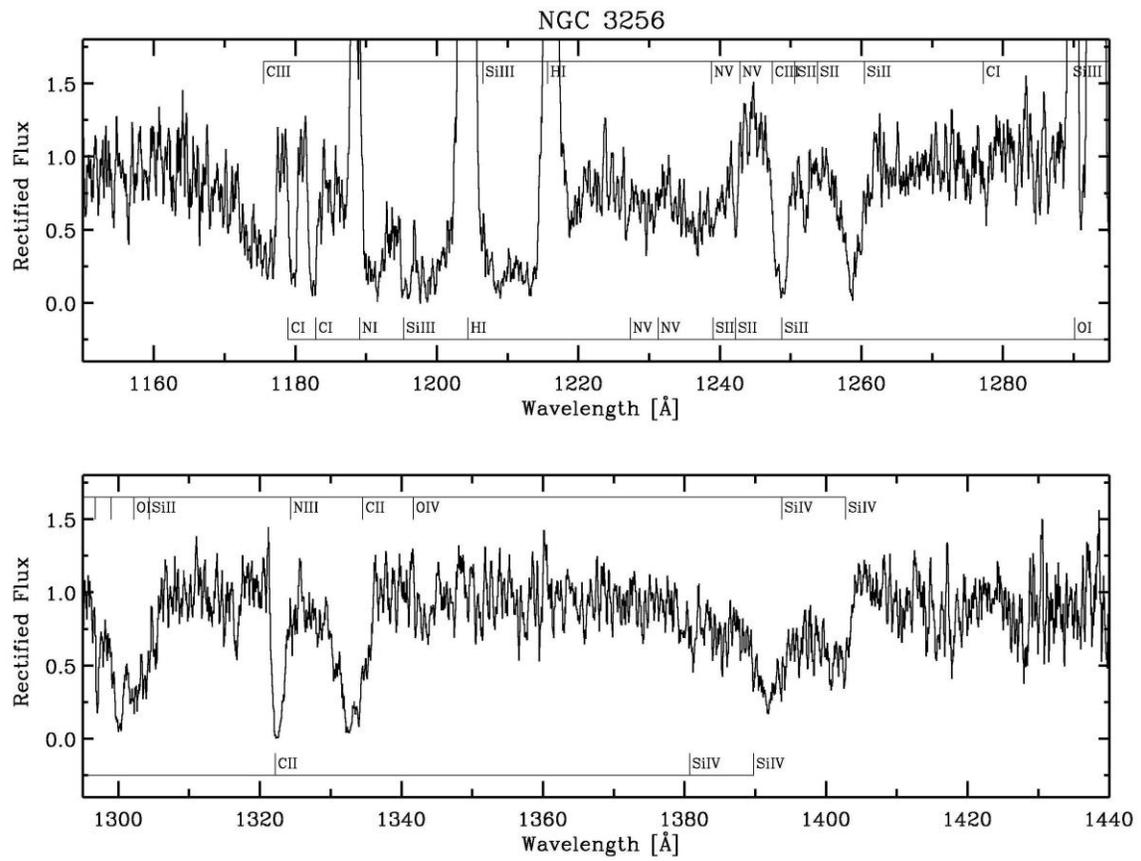

Figure 6. − Same as Figure 5, but for NGC 3256.

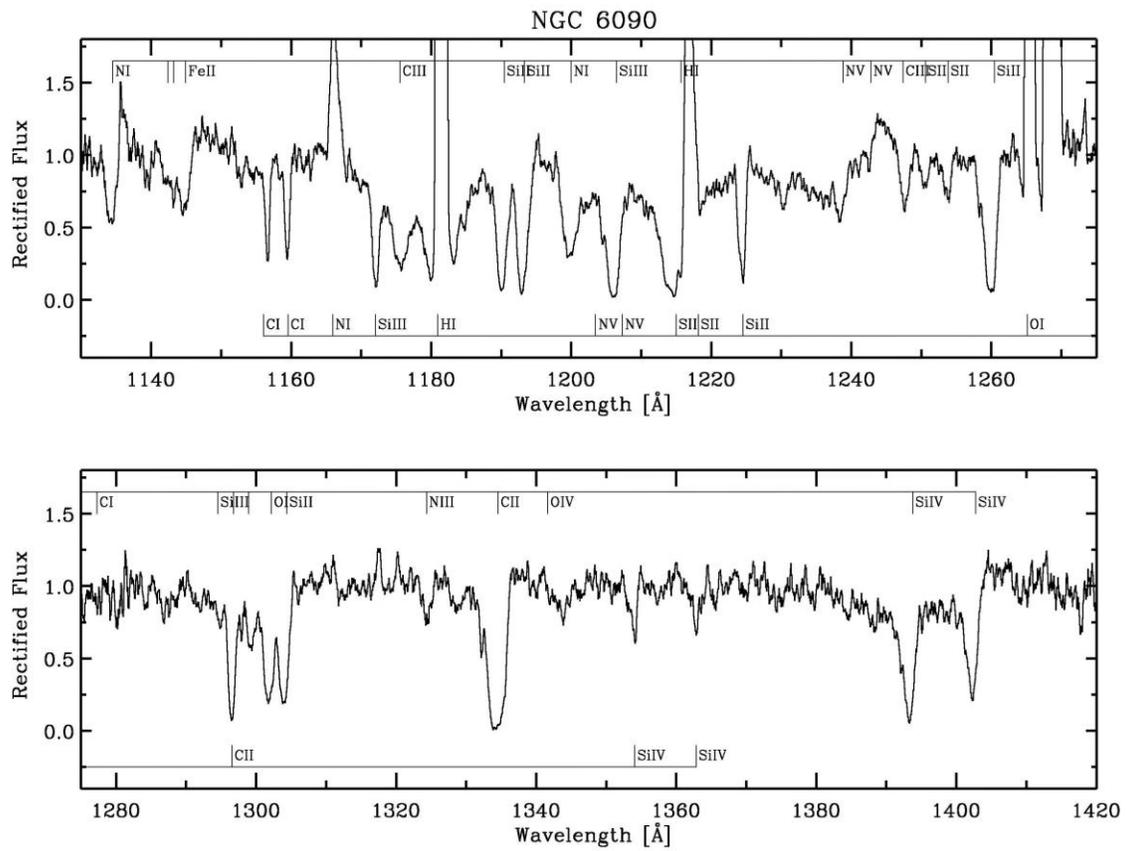

Figure 7. − Same as Figure 5, but for NGC 6090.

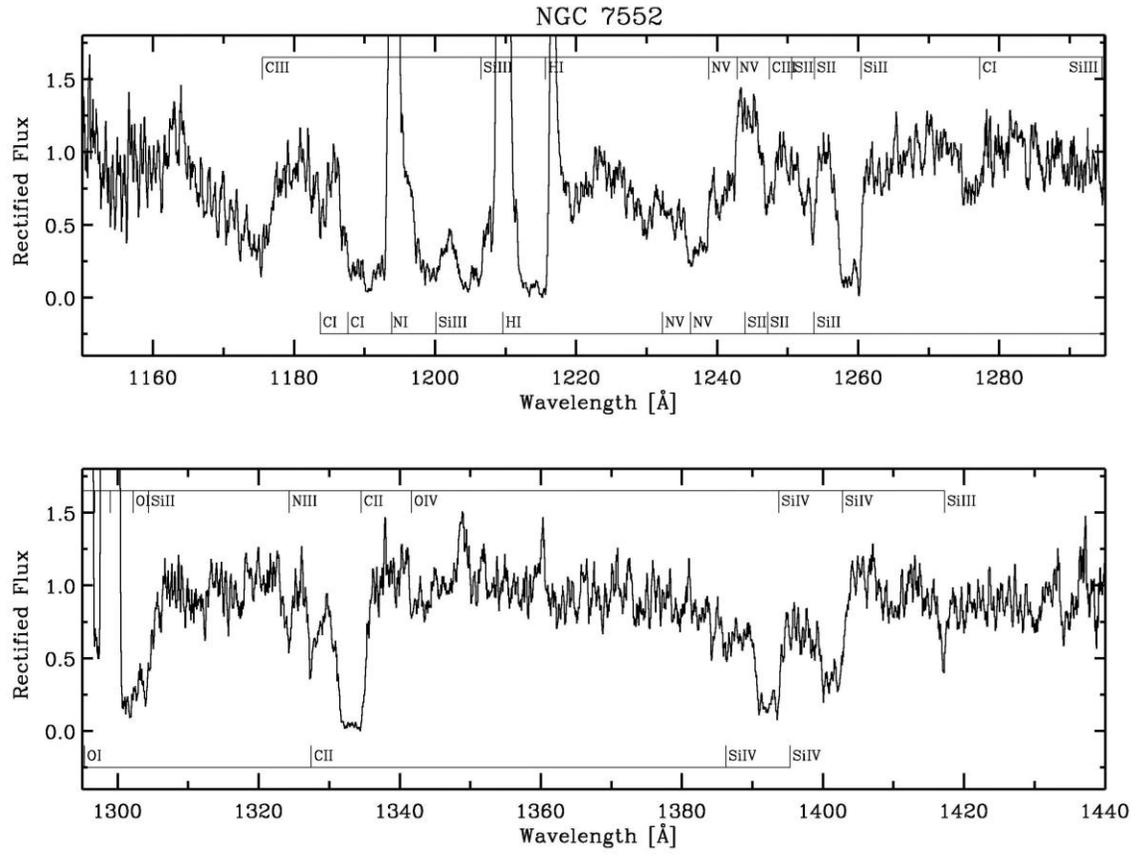

Figure 8. – Same as Figure 5, but for NGC 7552.

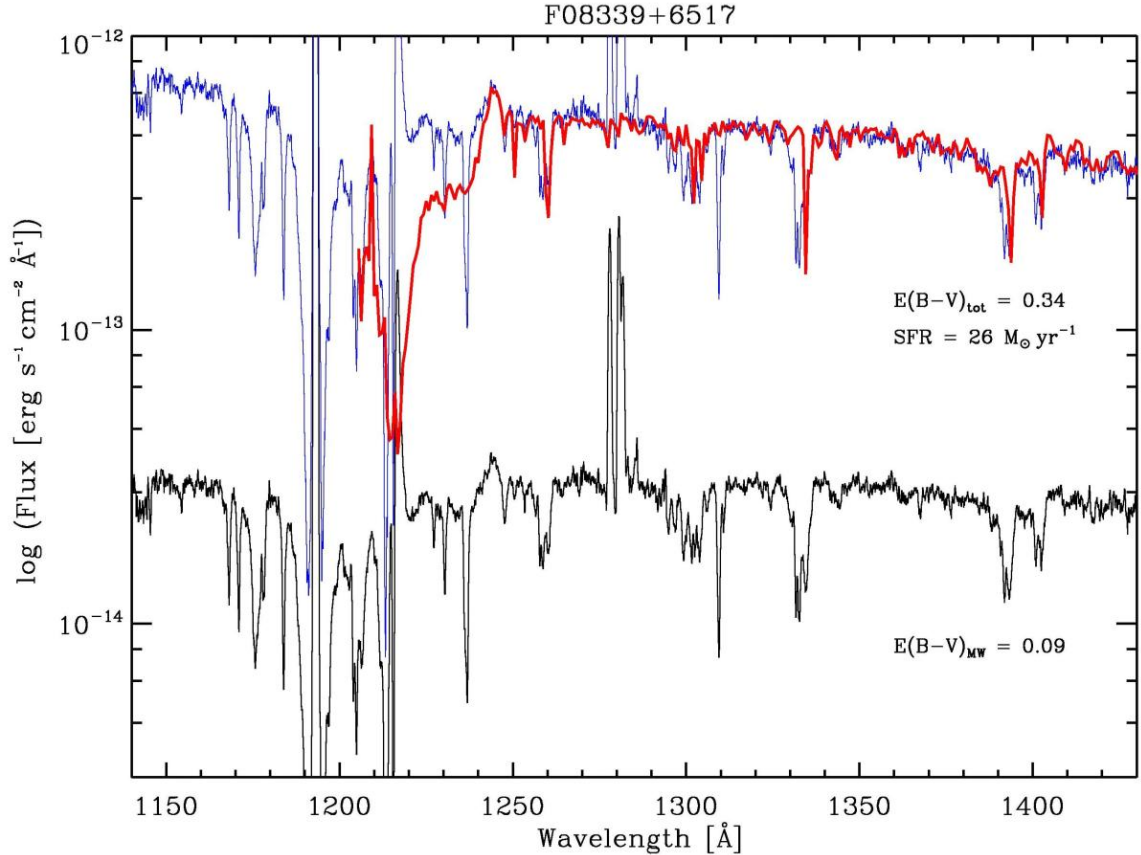

Figure 9. − Stellar population modeling of F08339+6517. The lower spectrum (black) are the observations processed through all steps described in Section 3, including the correction for foreground reddening $E(B-V)_{MW}$. The upper spectrum (blue) was produced by correcting the lower spectrum for the additional intrinsic dust reddening to obtain the total reddening $E(B-V)_{tot} = 0.34$. The superimposed spectrum (red) drawn with the thick line is a theoretical stellar spectrum with solar chemical composition, Kroupa IMF, and age 20 Myr. Note that the theoretical spectrum does not account for the interstellar lines. The corresponding star-formation rate is 26 $M_\odot$ yr$^{-1}$.

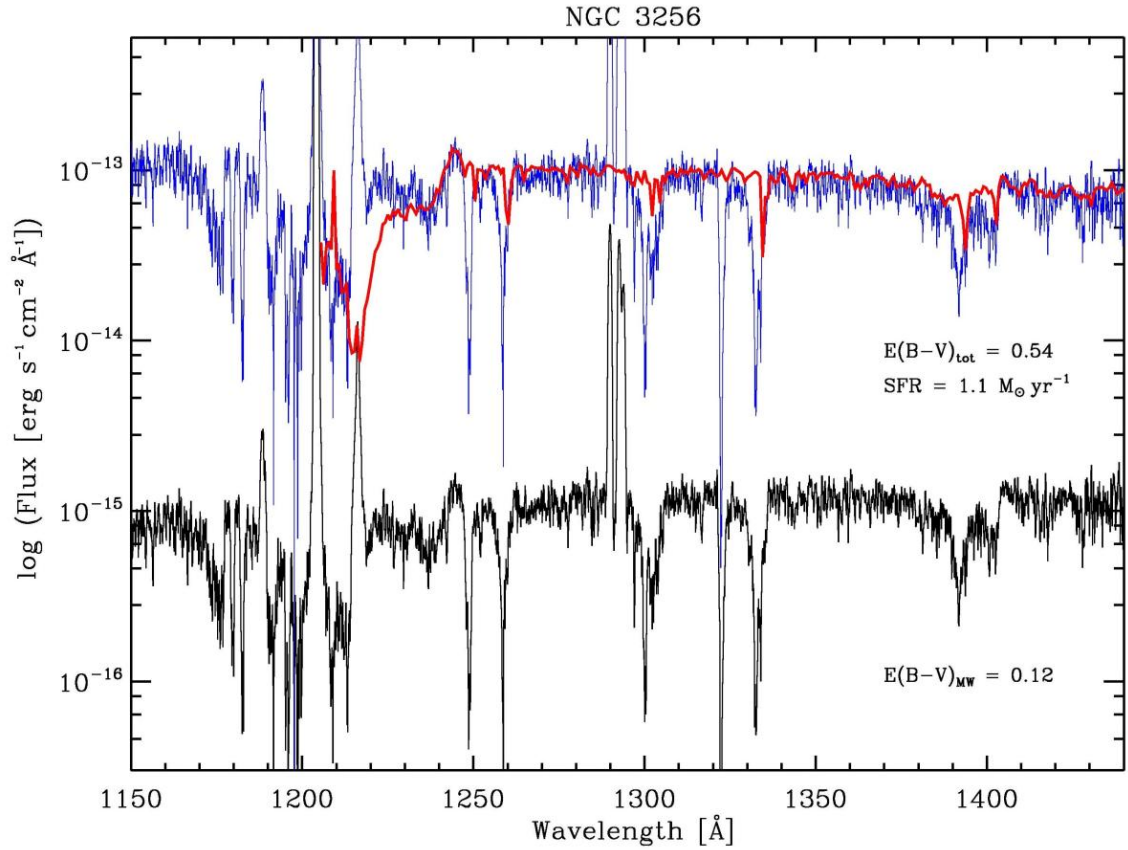

Figure 10. – Same as Figure 9, but for NGC 3256. The derived total reddening is $E(B-V)_{tot} = 0.54$, and the star-formation rate is $SFR = 1.1$ $M_\odot$ yr$^{-1}$.

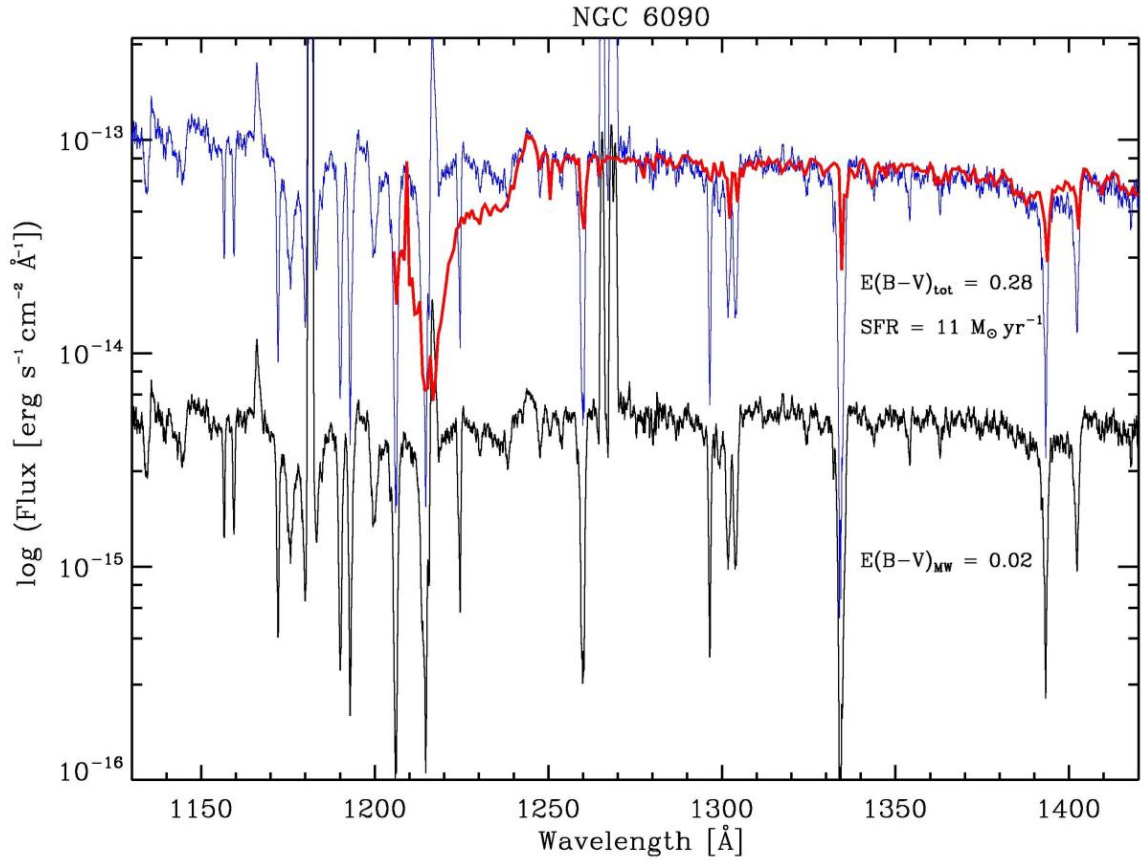

Figure 11. – Same as Figure 9, but for NGC 6090. The derived total reddening is

$E(B–V)_{tot}$ = 0.28, and the star-formation rate is $SFR$ = 11 $M_\odot$ yr$^{-1}$.

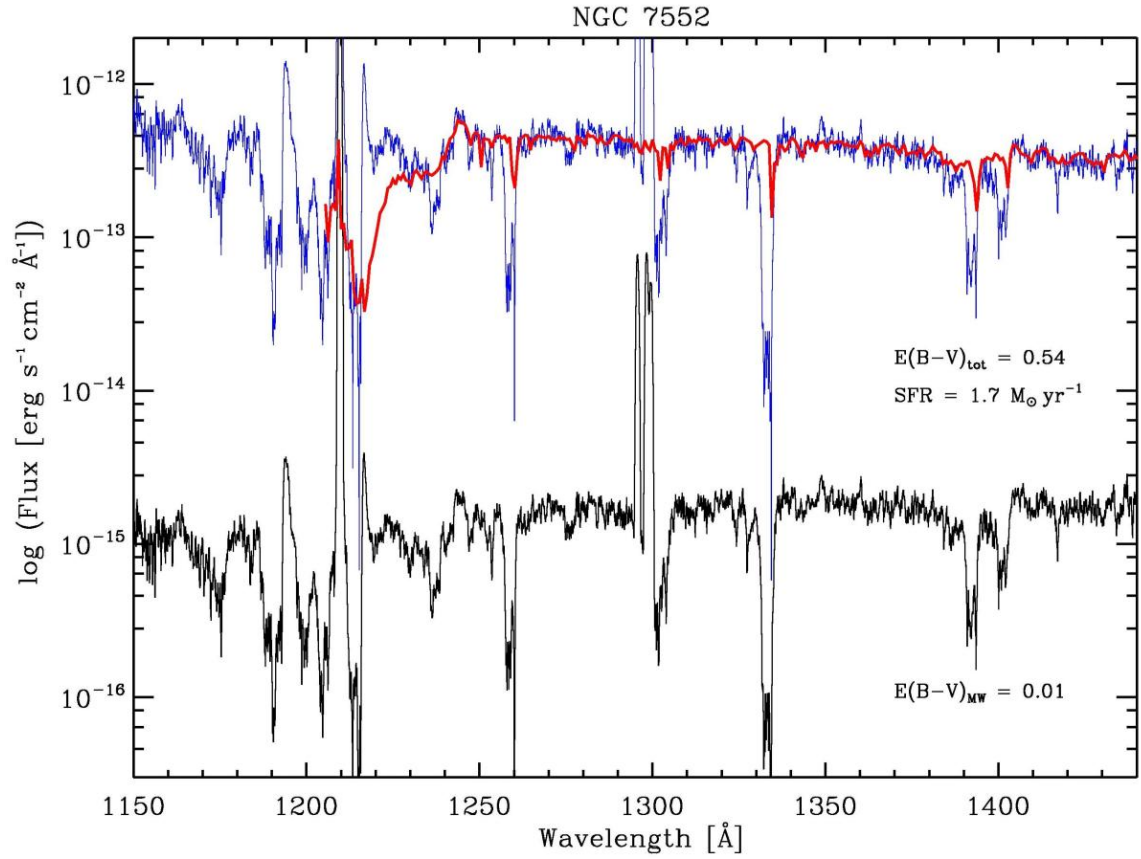

Figure 12. − Same as Figure 9, but for NGC 7552. The derived total reddening is $E(B-V)_{tot} = 0.54$, and the star-formation rate is $SFR = 1.7$ M$_\odot$ yr$^{-1}$.

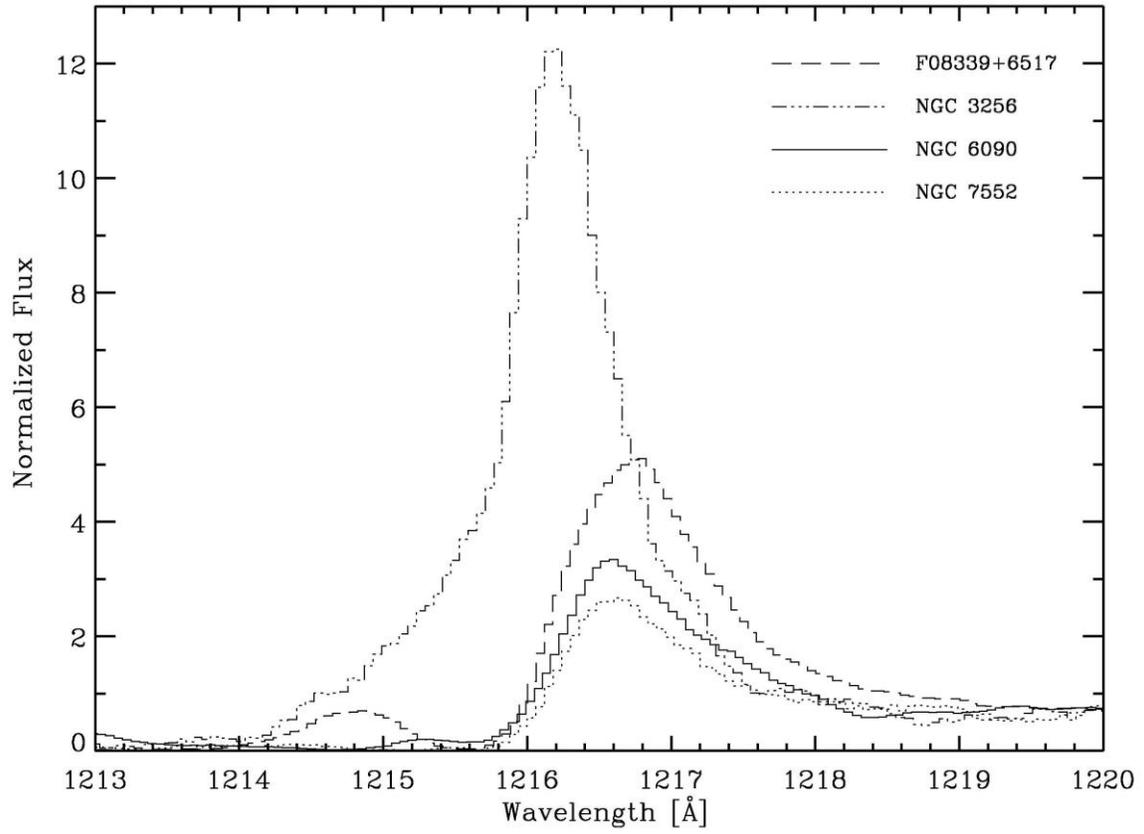

Figure 13. − Lyman-α profiles of the four galaxies. The dashed, dash-dotted, solid, and dotted lines indicate F08339+6517, NGC 3256, NGC 6090, and NGC 7552, respectively.

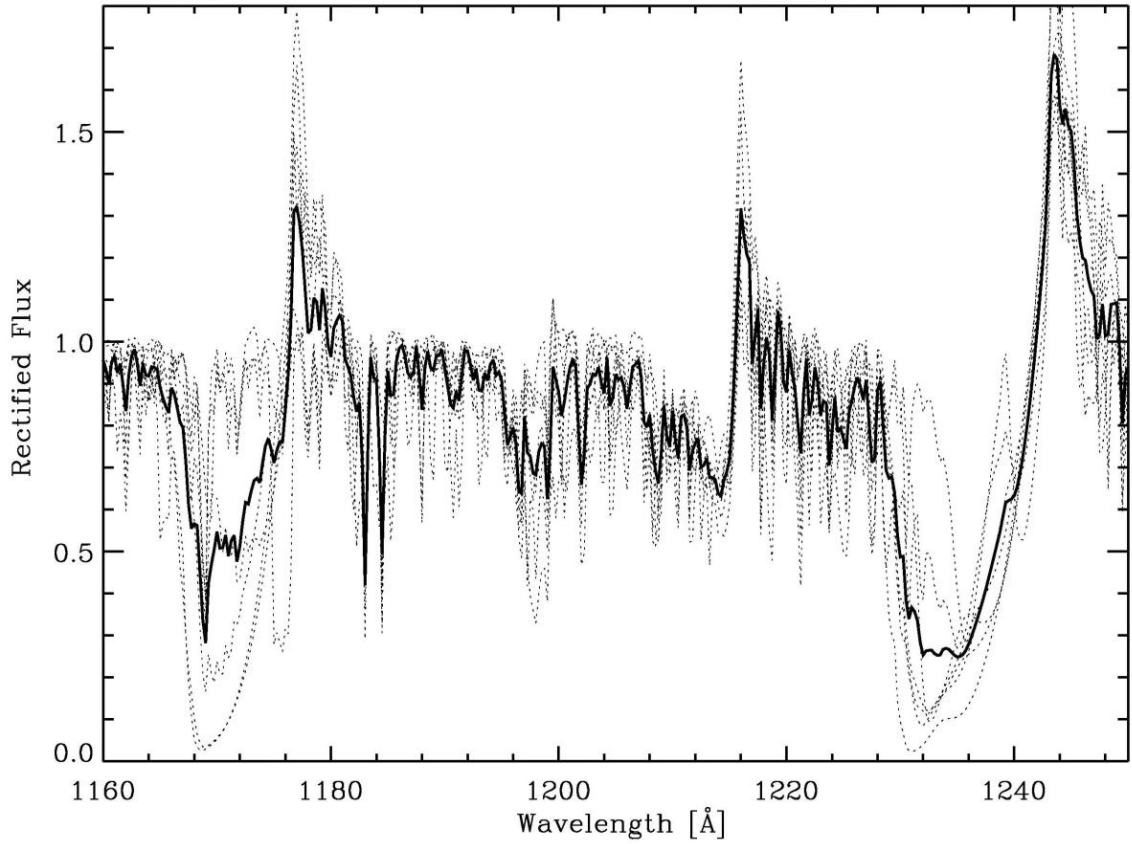

Figure 14. – Spectral region around Lyman-α in theoretical stellar spectra computed with CMFGEN. Dotted: individual spectra of HD 14947, HD 15570, HD 66811, HD 163958, HD 190429A, HD 192629, and HD 210839; solid: average of the seven individual spectra.

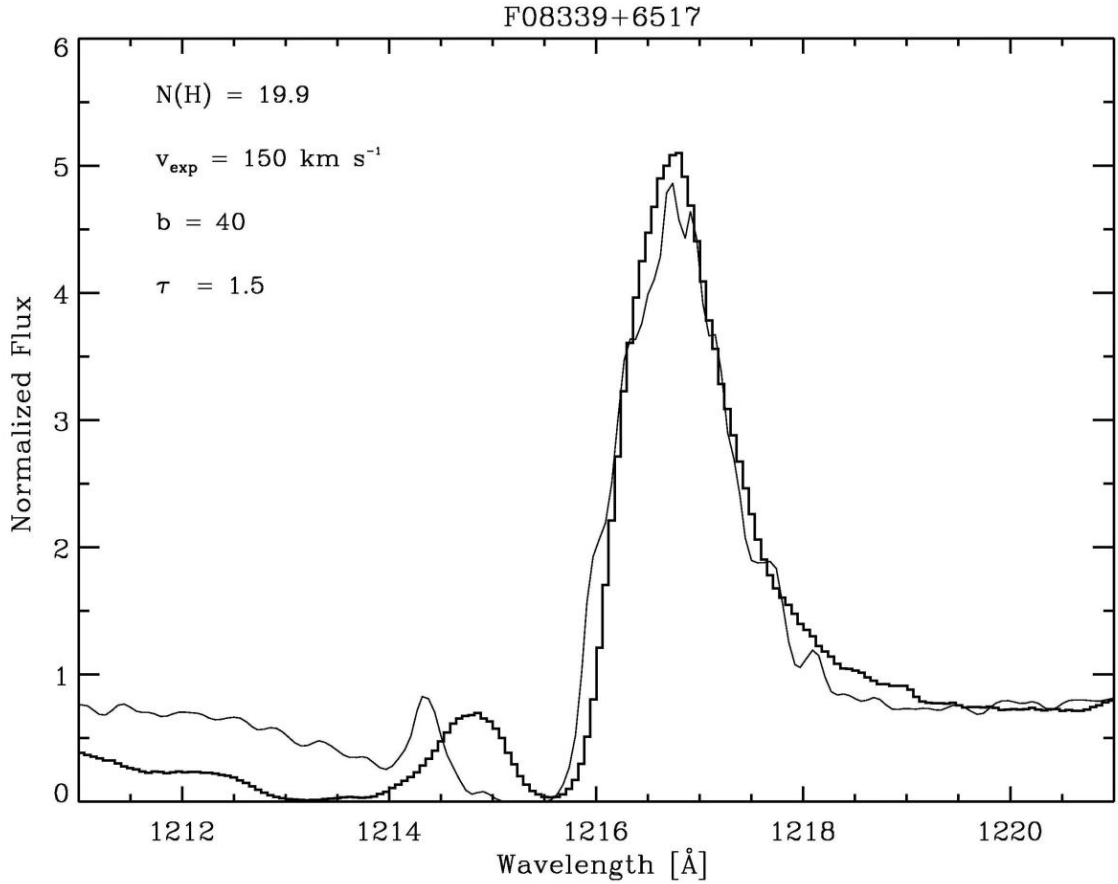

Figure 15. – Model fit to the observed Lyman-α profile of F08339+6517. The data and models are the thick and thin lines, respectively. The theoretical profile was computed with the 3-D radiation transfer code of Schaerer et al. (2011). The fit parameters are listed in the upper left corner (see text for details).

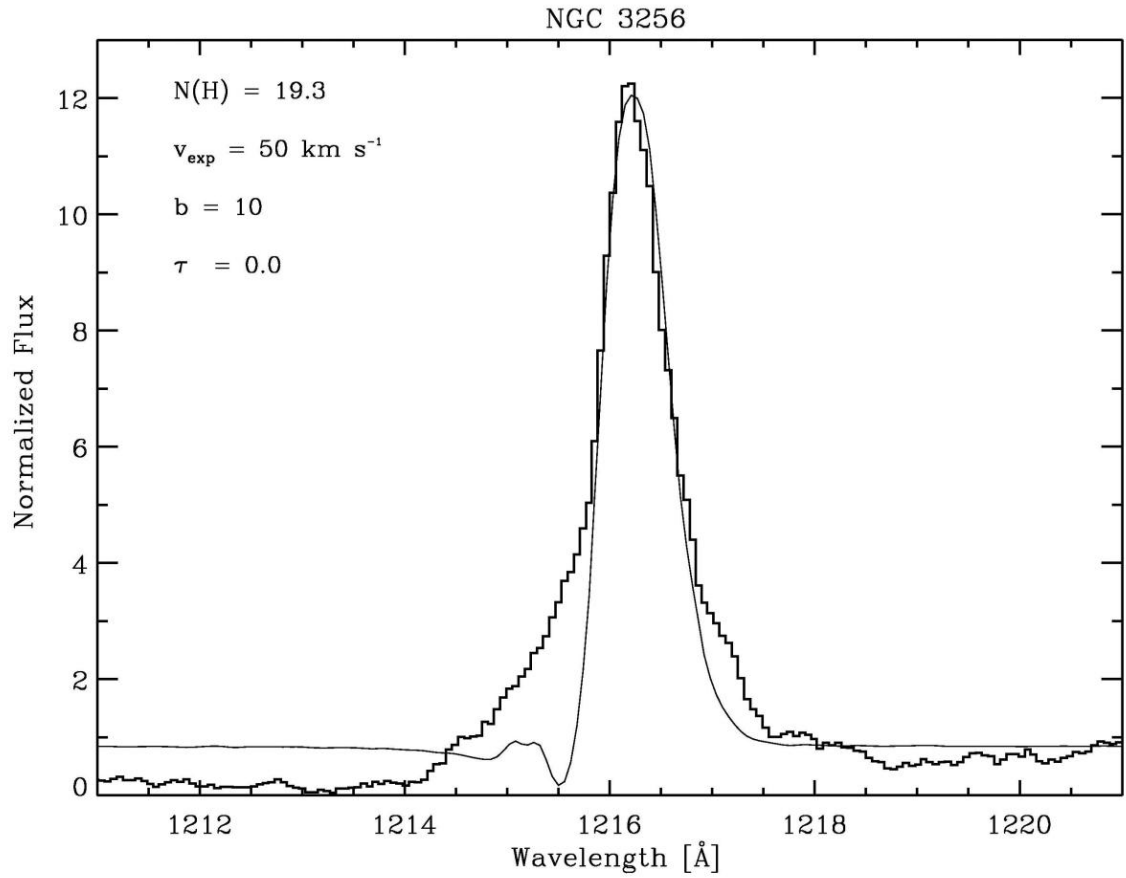

Figure 16. – Same as Figure 15 but for NGC 3256.

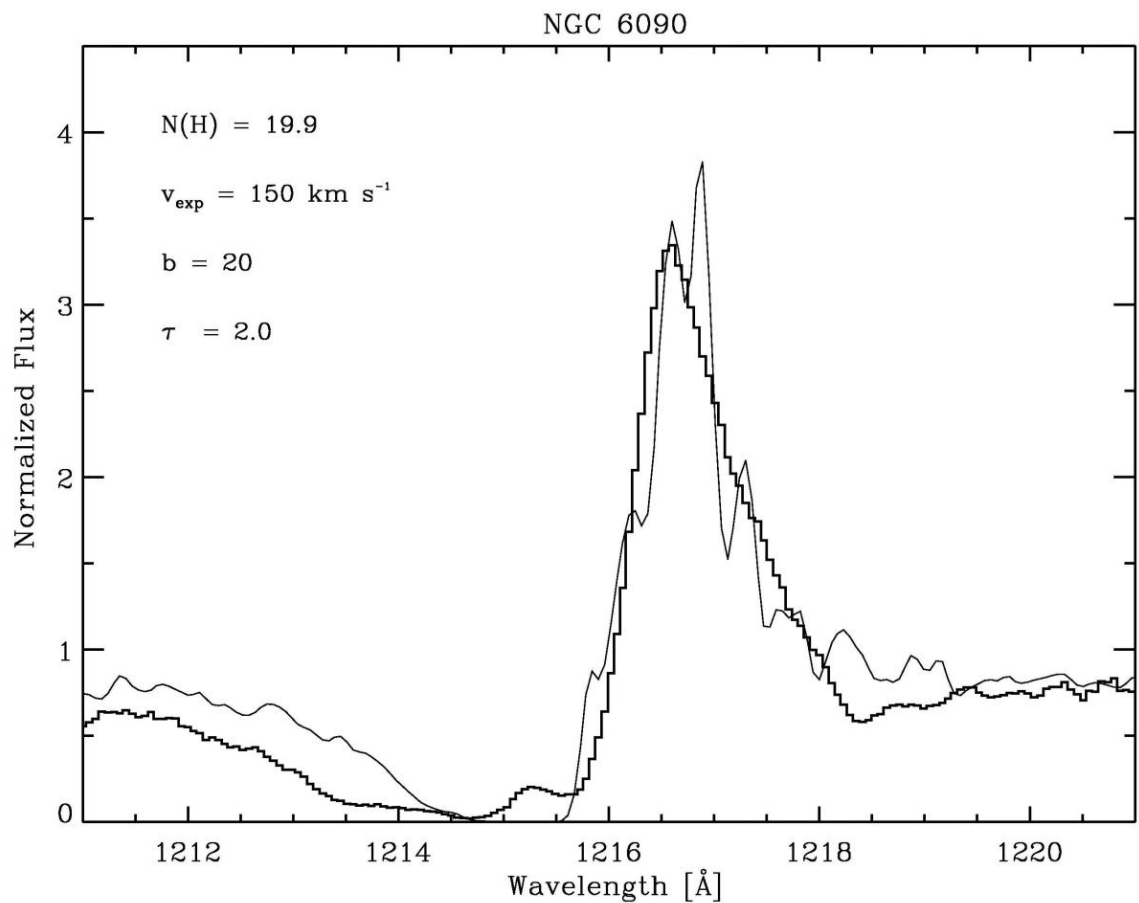

Figure 17. – Same as Figure 15 but for NGC 6090.

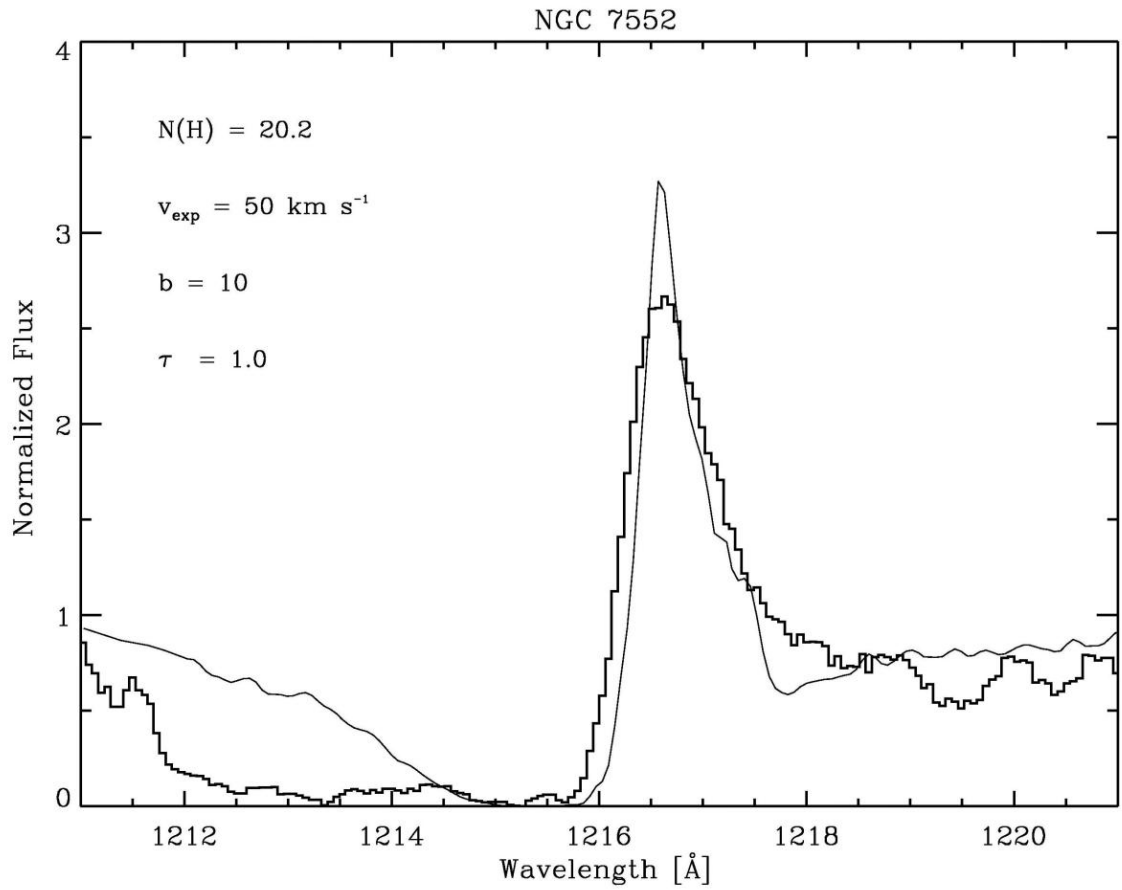

Figure 18. – Same as Figure 15 but for NGC 7552.

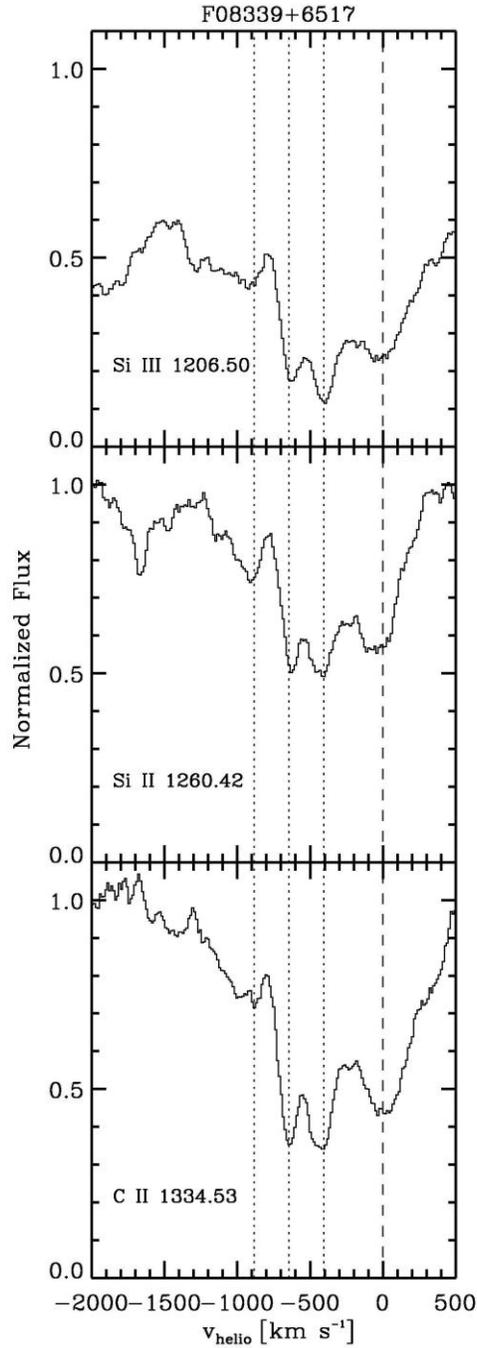

Figure 19. – Continuum-normalized low-ionization lines versus $v_{helio}$ for F08339+6517. The lines are from top to bottom: Si III λ1206, Si II λ1260, and CII λ1334. The dashed vertical line indicates zero velocity. Dotted vertical lines indicate discrete components at −406, −644, and −883 km s$^{-1}$.

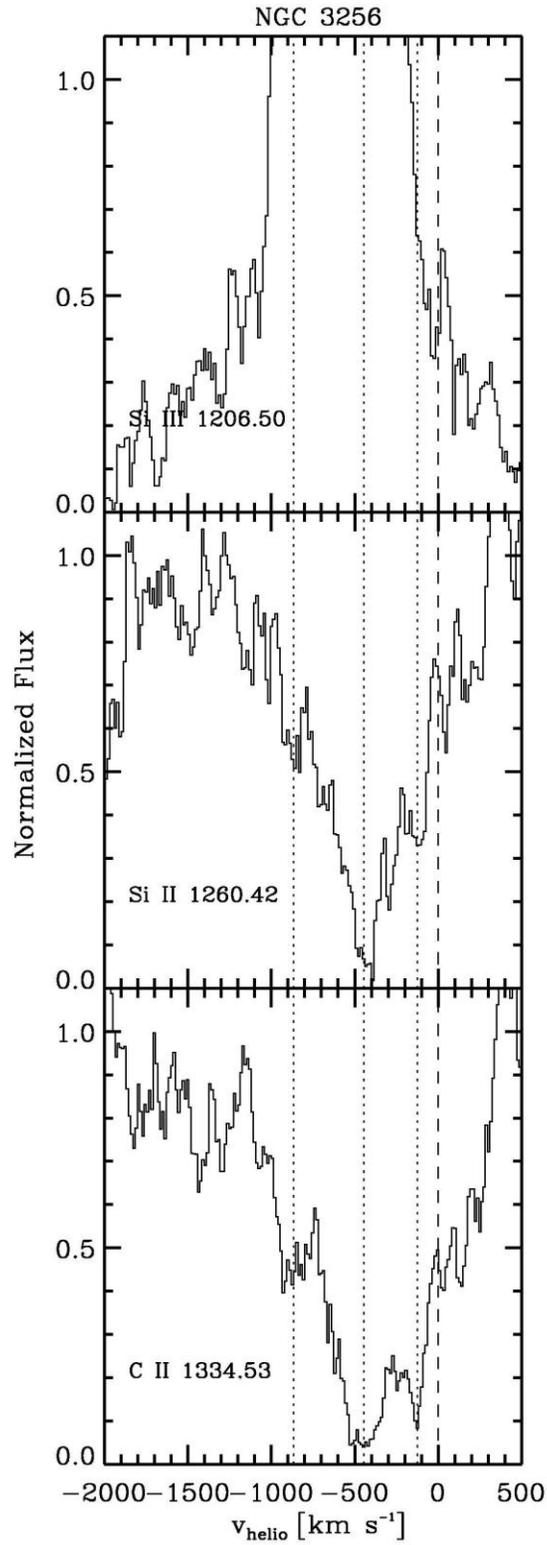

Figure 20. − Same as Figure 19, but for NGC 3256. The discrete components are at −126, −447, and −867 km s$^{-1}$.

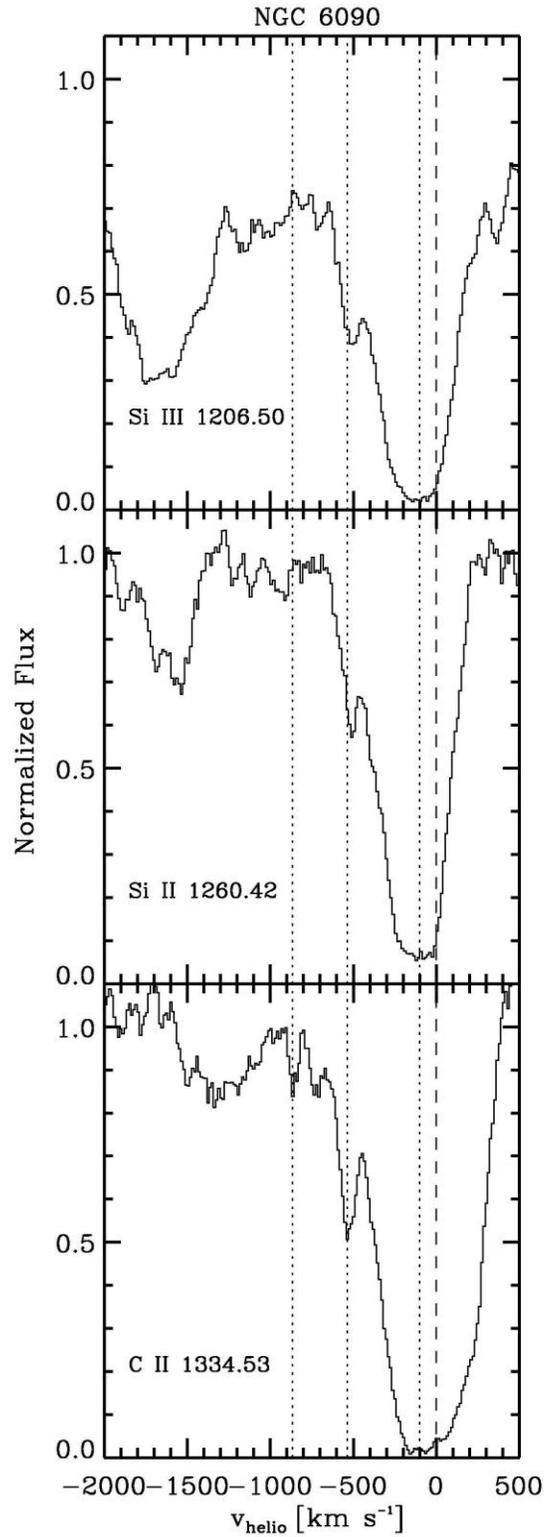

Figure 21. − Same as Figure 19, but for NGC 6090. The discrete components are at −101, −537, and −865 km s$^{-1}$.

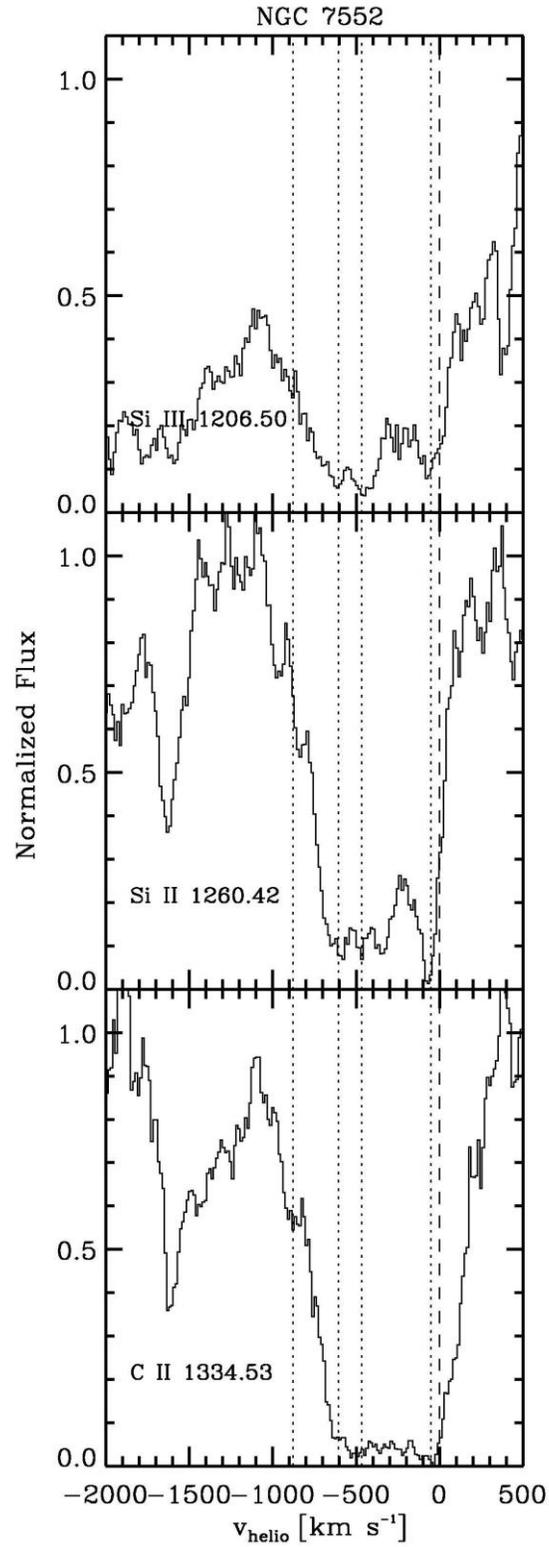

Figure 22. − Same as Figure 19, but for NGC 7552. The discrete components are at −52, −466, −605, and −879 km s$^{-1}$.

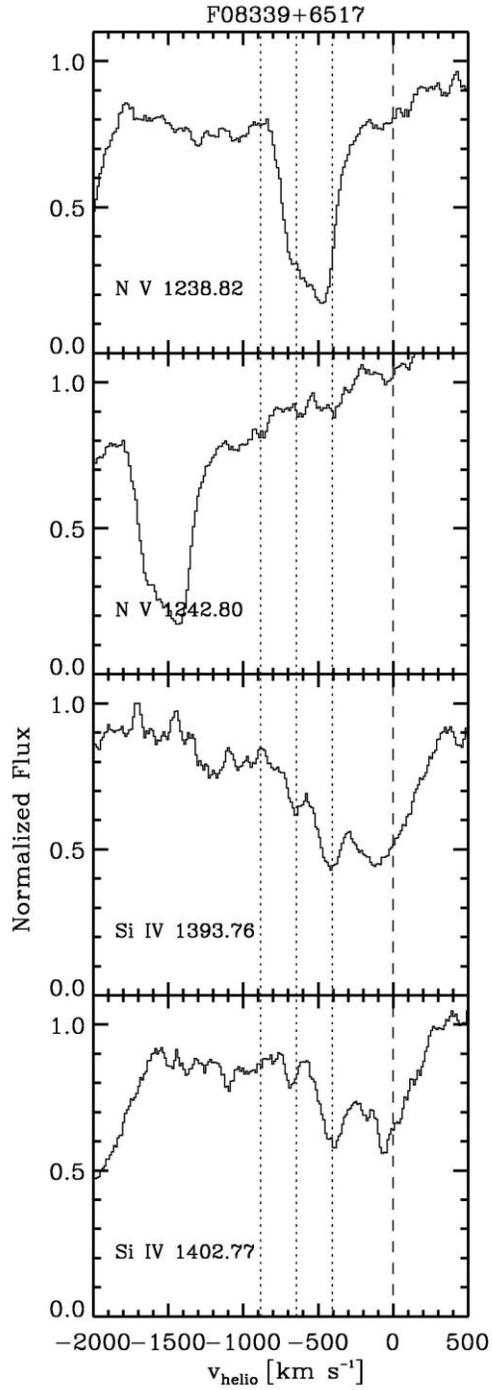

Figure 23. – Continuum-normalized high-ionization lines versus $v_{helio}$ for F08339+6517. The lines are from top to bottom: N V λ1238, N V λ1242, Si IV λ1393, and Si IV λ1402. The dashed vertical line indicates zero velocity. Dotted vertical lines indicate the velocities measured for the low-ionization lines in Figure 19.

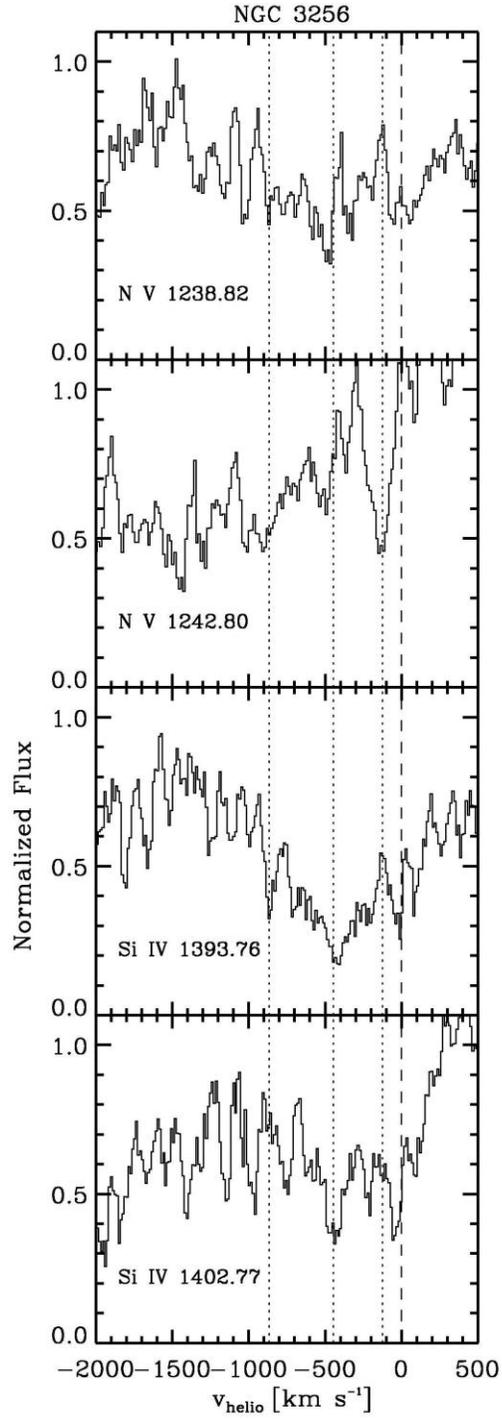

Figure 24. – Same as Figure 23, but for NGC 3256.

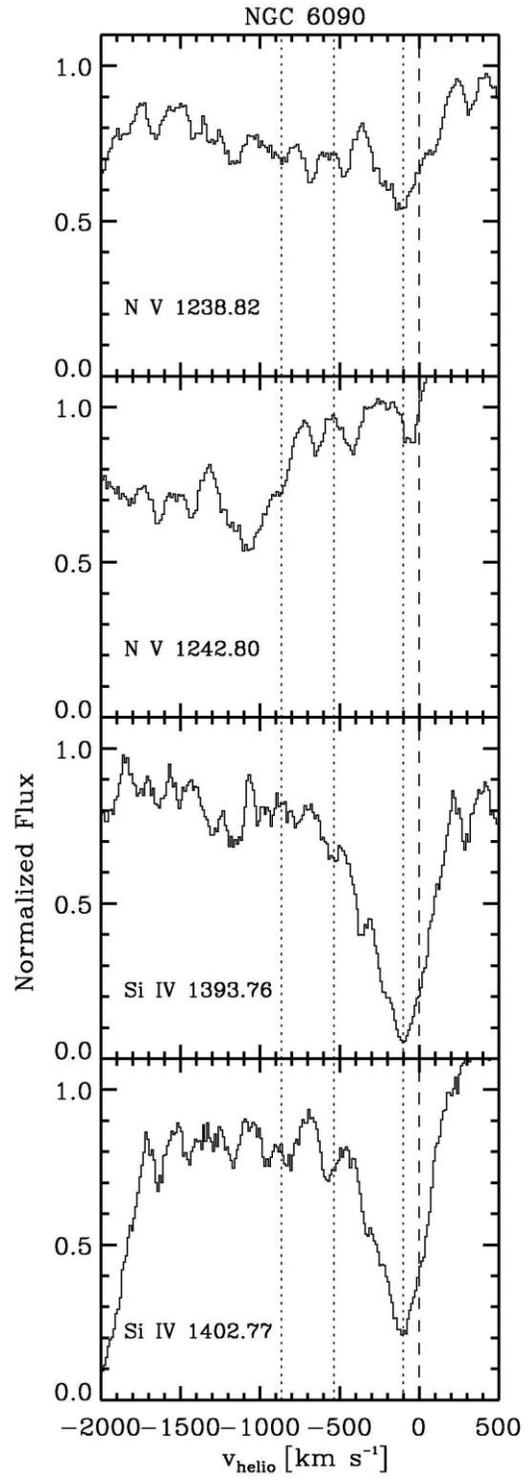

Figure 25. − Same as Figure 23, but for NGC 6090.

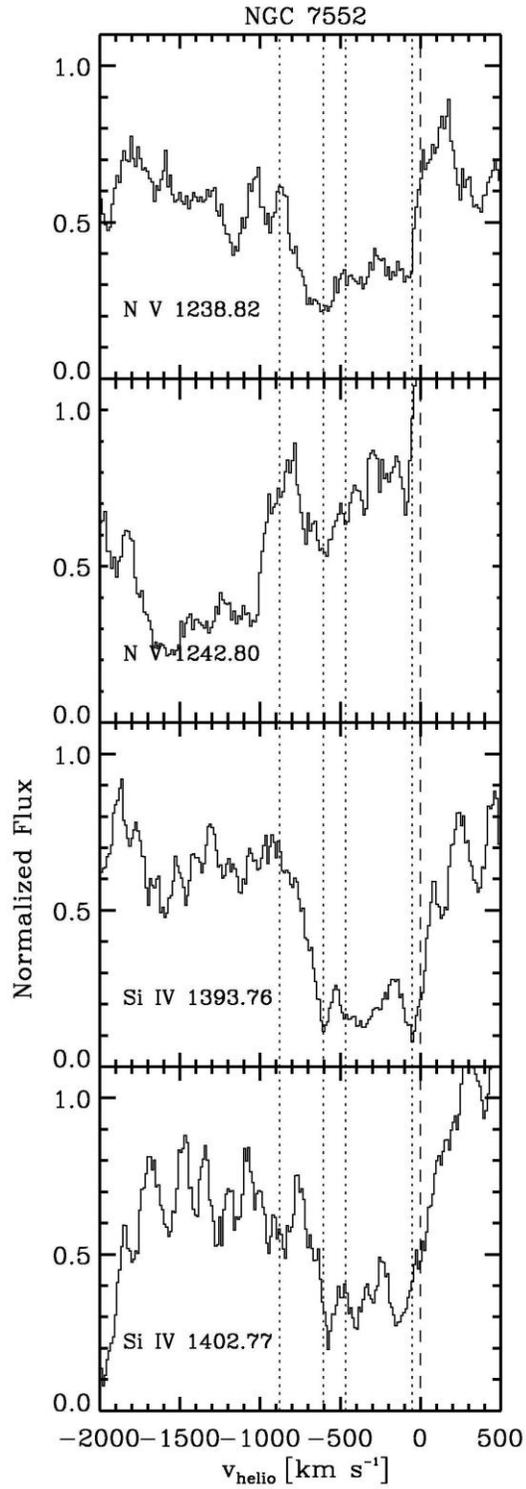

Figure 26. − Same as Figure 23, but for NGC 7552.

# Tables

Table 1. Galaxy properties

| Galaxy | Other Name | Morphology | $E(B-V)_{MW}$ | $v_{helio}$ (km s$^{-1}$) | $D$ (Mpc) | $A_{COS}$ (pc) | $\log(L_{IR})$ ($L_\odot$) | 12+log O/H |
|---|---|---|---|---|---|---|---|---|
| IRAS F08339+6517 | — | Pec | 0.092 | 5730 | 86.3 | 1048 | 11.11 | 8.45  (1) |
| IRAS F10257−4339 | NGC 3256 | Sb(s) pec | 0.121 | 2804 | 38.9 | 472 | 11.64 | 8.73  (2) |
| IRAS F16104+5235 | NGC 6090 | Sd pec | 0.020 | 8785 | 137 | 1658 | 11.58 | 8.77  (3) |
| IRAS F23133−4251 | NGC 7552 | SBbc | 0.014 | 1608 | 23.5 | 285 | 11.11 | 8.74  (4) |

References for 12+log O/H:
1) López-Sánchez et al. (2006)
2) Engelbracht et al. (2008)
3) Storchi-Bergmann, Calzetti, & Kinney (1994)
4) Calzetti et al. (2010)

Table 2. COS observations of IR-bright galaxies

| HST ID | Galaxy | Start Time | Central Wavelength (Å) | Wavelength Range (Å) | Duration (s) |
|---|---|---|---|---|---|
| LBII01010 | F08339+6517 | 2011-05-13 21:19:04 | 1309 | 1137 – 1432 | 1303 |
| LBII01020 | F08339+6517 | 2011-05-13 21:44:33 | 1327 | 1157 – 1449 | 1303 |
| LBII02010 | NGC 3256 | 2011-06-19 08:05:21 | 1291 | 1130 – 1428 | 2553 |
| LBII02020 | NGC 3256 | 2011-06-19 09:30:44 | 1300 | 1139 – 1437 | 3055 |
| LBII02030 | NGC 3256 | 2011-06-19 11:06:33 | 1309 | 1148 – 1445 | 3055 |
| LBII02040 | NGC 3256 | 2011-06-19 12:42:23 | 1318 | 1157 – 1454 | 3055 |
| LBII03010 | NGC 6090 | 2011-10-21 19:19:00 | 1309 | 1126 – 1417 | 2588 |
| LBII03020 | NGC 6090 | 2011-10-21 20:42:28 | 1318 | 1135 – 1426 | 3159 |
| LBII03030 | NGC 6090 | 2011-10-21 22:18:17 | 1327 | 1143 – 1435 | 3159 |
| LBII04010 | NGC 7552 | 2011-05-10 21:31:58 | 1300 | 1144 – 1442 | 2544 |
| LBII04020 | NGC 7552 | 2011-05-10 22:57:40 | 1318 | 1162 – 1469 | 3055 |

Table 3. Star-formation properties derived from the UV

| Galaxy | $E(B-V)_{int}$ | $\log(L_{COS})$ (erg s$^{-1}$ Å$^{-1}$) | $SFR_{COS}$ (M$_\odot$ yr$^{-1}$) | $SFR_{IUE}$ (M$_\odot$ yr$^{-1}$) |
|---|---|---|---|---|
| F08339+6517 | 0.25 | 41.6 | 26 | 70 |
| NGC 3256 | 0.42 | 40.2 | 1.1 | 55 |
| NGC 6090 | 0.26 | 41.2 | 11 | 29 |
| NGC 7552 | 0.53 | 40.4 | 1.7 | 57 |

$SFR_{COS}$ and $SFR_{IUE}$ are derived from the UV luminosities in the COS and IUE apertures. The circular COS aperture has 2.5″ diameter, and the rectangular IUE aperture has 10″ by 20″ dimensions.

Table 4. Rotation velocities of the galaxies

| Galaxy | $W_{20}$ (km s$^{-1}$) | Reference | $W_R$ (km s$^{-1}$) | $i$ (°) | $W_{rot}$ (km s$^{-1}$) |
|---|---|---|---|---|---|
| F08339+6517 | 338 | 1 | 300 | 13 | (339) |
| NGC 3256 | 306 | 2 | 268 | 18 | 434 |
| NGC 6090 | 357 | 3 | 319 | 27 | 351 |
| NGC 7552 | 257 | 2 | 219 | 28 | 233 |

References for $W_{20}$:
1) Martin et al. (1991)
2) Koribalski et al. (2004)
3) van Driel et al. (2001)